\documentclass{article}

\usepackage{microtype}
\usepackage{graphicx}
\usepackage{subfigure}
\usepackage{booktabs} % for professional tables
\usepackage[numbers]{natbib}

\usepackage{hyperref}
\usepackage{xcolor, colortbl}         % colors
\definecolor{mygray}{gray}{0.6}
\definecolor{mygray-bg}{gray}{0.9}

\usepackage{enumitem}

\usepackage[accepted]{icml2024}

\usepackage{amsmath}
\usepackage{amssymb}
\usepackage{mathtools}
\usepackage{amsthm}

\usepackage[capitalize,noabbrev]{cleveref}

\theoremstyle{plain}

\theoremstyle{definition}

\theoremstyle{remark}

\usepackage[textsize=tiny]{todonotes}

\begin{document}

\twocolumn[
\icmltitle{From Vision to Audio and Beyond: A Unified Model for Audio-Visual Representation and Generation}



\icmlsetsymbol{equal}{*}

\begin{icmlauthorlist}
\icmlauthor{Kun Su}{yyy,equal}
\icmlauthor{Xiulong Liu}{yyy,equal}
\icmlauthor{Eli Shlizerman}{yyy,sch}
\end{icmlauthorlist}

\icmlaffiliation{yyy}{Department of ECE, University of Washington, Seattle, United States}
\icmlaffiliation{sch}{Department of Applied Math, University of Washington, Seattle, United States}

\icmlcorrespondingauthor{Eli Shlizerman}{shlizee@uw.edu}

\icmlkeywords{Machine Learning, ICML}

\vskip 0.3in
]



\printAffiliationsAndNotice{\icmlEqualContribution} % otherwise use the standard text.

\begin{abstract}
% Abstract goes here.
Video encompasses both visual and auditory data, creating a perceptually rich experience where these two modalities complement each other. As such, videos are a valuable type of media for the investigation of the interplay between audio and visual elements. Previous studies of audio-visual modalities primarily focused on either audio-visual representation learning or generative modeling of a modality conditioned on the other, creating a disconnect between these two branches. A unified framework that learns representation and generates modalities has not been developed yet. In this work, we introduce a novel framework called Vision to Audio and Beyond (VAB) to bridge the gap between audio-visual representation learning and vision-to-audio generation. The key approach of VAB is that rather than working with raw video frames and audio data, VAB performs representation learning and generative modeling within latent spaces. In particular, VAB uses a pre-trained audio tokenizer and an image encoder to obtain audio tokens and visual features, respectively. It then performs the pre-training task of visual-conditioned masked audio token prediction. This training strategy enables the model to engage in contextual learning and simultaneous video-to-audio generation. After the pre-training phase, VAB employs the iterative-decoding approach to rapidly generate audio tokens conditioned on visual features. Since VAB is a unified model, its backbone can be fine-tuned for various audio-visual downstream tasks. Our experiments showcase the efficiency of VAB in producing high-quality audio from video, and its capability to acquire semantic audio-visual features, leading to competitive results in audio-visual retrieval and classification.
\end{abstract}

\begin{figure}[tp]
    \centering
    \includegraphics[width=\linewidth]{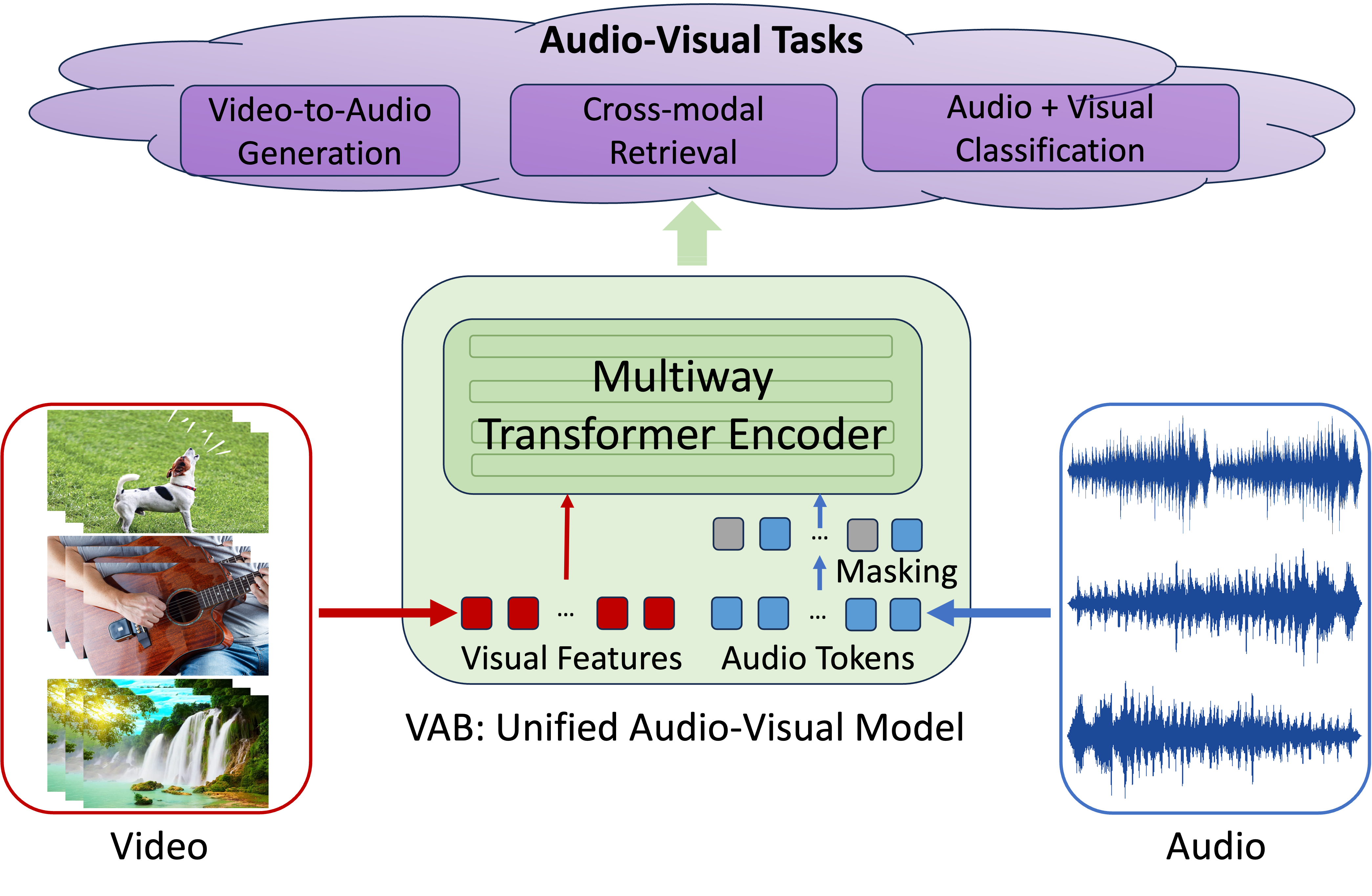}
    \caption{VAB is a unified audio-visual model capable of supporting various audio-visual tasks within a single framework. 
    }
    \label{fig:teaser}
\end{figure}

\section{Introduction}
\label{sec:intro}

% Audio-Visual
When presented with a video which audio has been muted, we have the ability to mentally hallucinate the corresponding sound based on our knowledge. This mental process encompasses the association of audio with visual elements and the creative aspect of sound imagination. In the field of artificial intelligence, we have witnessed the emergence of foundational models in the domains of vision and language~\cite{wang2022image, yu2022coca, alayrac2022flamingo} which are capable of supporting both the association and generation of content. In the realm of audio and vision, however, it remains relatively unexplored whether similar models can be proposed.

% Challenges
Creating a general-purpose audio-visual model poses significant challenges. First, raw video and audio signals are both high-dimensional and time-dependent data, creating an intricate scenario for comprehension of their joint events and necessitating extensive training computations. Existing contrastive learning and masked autoencoder-based audio-visual learners~\cite{gong2022contrastive, huang2022mavil} are deterministic models that cannot support generative purposes, and the reconstruction loss is made on the level of image and spectrogram patches. The state-of-the-art audio generative frameworks~\cite{kreuk2022audiogen, copet2023simple, mei2023foleygen}, on the other hand, usually perform modeling on the latent space and sophisticated models are often essential to generate high-quality audio, requiring multi-stage modeling~\cite{dhariwal2020jukebox, garcia2023vampnet, agostinelli2023musiclm}. It remains uncertain whether we can employ similar self-supervised representation learning techniques in the latent space while consistently producing high-quality audio.

% Ours
To address these challenges, we introduce an innovative audio-visual framework `Vision to Audio and Beyond' (VAB), which is an efficient unified audio-visual framework capable to learn to associate audio with visual signals and enables vision-to-audio generation within the same model. At its core, this framework involves a pre-training task of predicting masked audio from visual inputs. In order to facilitate the learning of audio-visual representation and audio generation, we perform the pre-training task within the latent space instead of using raw images and audio spectrograms. Specifically, we tokenize the audio data into discrete tokens utilizing a public open pre-trained neural audio codec and extract frame-level visual features from a self-supervised pre-trained image encoder. During VAB pre-training, we employ an encoder-only multi-way transformer to predict masked audio tokens from visual features using a variable masking scheme. After completion of the pre-training phase, the VAB model can function as a uni-modal or multi-modal encoder, and it is prepared for fine-tuning in cross-modal retrieval and classification tasks. Meanwhile, it can also support zero-shot visual-conditioned audio generation through efficient parallel decoding strategies.

We conduct comprehensive experiments across a range of downstream tasks to demonstrate our approach and its advantages. These tasks include vision-to-audio generation (on VGGSound), audio-visual event classification (on AudioSet-20K, AudioSet-2M, VGGSound), audio-to-video and video-to-audio retrieval (on AudioSet, VGGSound, MSR-VTT), and audio-only classification (on ESC-50, SPC). Our experimental results showcase that the VAB model can efficiently produce high-fidelity audio from silent video, achieving a speedup of 17 times than the state-of-the-art autoregressive approach. Furthermore, the representations learned from VAB yield competitive outcomes across various classification and retrieval tasks.
% Contribution
In summary, our contributions are as follows:
\begin{itemize}[align=right,itemindent=0em,labelsep=2pt,labelwidth=1em,leftmargin=*,itemsep=-0.5em]
    \item We propose a first-of-its-kind audio-visual foundation and unified framework, named VAB, that supports both audio-visual representation learning and visual-conditioned audio generation in the latent space.
    \item VAB model can efficiently generate 10 seconds of high-quality audio from visual inputs with the power of parallel decoding. To the best of our knowledge, this is the first employment of parallel decoding on the task of vision-to-audio generation.
    \item Comprehensive experiments demonstrate that the foundation VAB model could be fine-tuned for various audio-visual downstream tasks and achieve state-of-the-art performances in audio-visual retrieval, audio-visual event classification, and audio-only classification.
\end{itemize}

\section{Related Work}
\label{sec:related}

\subsection{Video to Audio Generation}
Generating audio given video stream has garnered growing interest, in particular as generative models have advanced rapidly in recent years in both visual and audio domains. Efforts exploring various facets of the connections between audio and videos have emerged. Earlier works proposed to leverage the semantic alignment between audio and videos with the goal of generating natural sounds from objects or scenes depicted in the video~\cite{owens2016visually, mehri2016samplernn, zhou2018visual}. Subsequently, several studies have focused on the generation of music from videos, capitalizing on both temporal and semantic correlations between music signal characteristics and human body movements~\cite{su2020audeo, gan2020foley, su2020multi, su2021does, Liu_2024_WACV}. More recently, we have seen the emergence of more versatile video-to-music~\cite{di2021video, su2023v2meow} or video-to-audio~\cite{iashin2021taming, sheffer2023hear, mei2023foleygen} generation methods designed for in-the-wild videos. These approaches are based on generative modeling within discrete tokenized audio and music signals to yield higher-fidelity audio outputs. In parallel, diffusion-based vision-to-audio modeling has been introduced as well~\cite{su2023physics, luo2023diff}. Such generative models are also designed to specialize for generation and are not suitable for broader audio-visual understanding and tasks. In our work, our objective is not only to achieve high-quality audio generation from video but also to facilitate cross-modal representation learning, i.e. a foundation and unified model, encompassing both audio and visual modalities.

\subsection{Audio-visual Representation Learning}
The process of learning representations for audio and visual data has been explored through both supervised and self-supervised approaches. In earlier studies, deep neural networks were trained on annotated audio-visual pairs using supervised methods~\cite{ngiam2011multimodal, kim2013deep, nagrani2021attention, liu2024tackling}. Subsequently, self-supervised learning approaches have harnessed the correspondence between audio and visual data to acquire representations~\cite{aytar2016soundnet,arandjelovic2017look,arandjelovic2018objects}, which are then employed in various downstream tasks such as audio-visual classification and retrieval~\cite{chen2020vggsound, gemmeke2017audio, Dong_2024_CVPR}. Among these methods, contrastive learning and masked autoencoder based approaches applied to audio and visual signals have emerged as effective learning paradigms~\cite{ma2020active,morgado2021audio,gong2022contrastive,huang2022mavil}. Audio-visual contrastive learning capitalizes on the similarity between audio-visual pairs within the same video and across different videos, using them as self-supervised signals to define a discriminative objective~\cite{ma2020active,morgado2021audio}. In contrast, masked autoencoder methods draw their inspiration from context-aware learning in natural language processing, e.g., BRNN~\cite{NIPS2015_c75b6f11} and BERT model~\cite{devlin2019bert}, and have since been extended to encompass other signals types, including skeleton sequences~\cite{Su_2020_CVPR}, images~\cite{he2022masked}, videos~\cite{tong2022videomae}, audio~\cite{huang2023masked}. The combination of these two paradigms demonstrates cumulative improvements in audio-visual representations~\cite{gong2022contrastive,huang2022mavil, georgescu2023audiovisual}. Current audio-visual representation learning mainly utilizes raw video frames and audio spectrogram data as input, aiming to incorporate maximal available information. However, it presents challenges in modeling, due to the high dimensionality and intricate patterns of the raw data. In this work, we seek to explore an alternative approach by leveraging latent audio tokens to enable both representation learning and generative modeling.

\subsection{Masked Generative Modeling}
Multiple works have incorporated masked generative modeling within their frameworks. MaskGIT~\cite{chang2022maskgit} proposed to utilize a bidirectional transformer for token modeling and introduced parallel decoding which enhances inference speed for image generation. MAGE~\cite{li2023mage} unified image generation and representation learning through a masking-based approach. Subsequently, Muse~\cite{chang2023muse} extended MaskGIT's capabilities to text-to-image generation. In addition, Magvit~\cite{yu2023magvit} leveraged a multi-task learning paradigm to efficiently generate video. In the realm of audio, Vampnet~\cite{garcia2023vampnet} employed a two-stage approach to masked acoustic token modeling, achieving unconditional music generation. Furthermore, SoundStorm~\cite{borsos2023soundstorm} showcased its capability to rapidly generate high-quality speech from text. Compared to these prior studies, the aim of this study is to unify audio and vision as vision-to-audio generation and various audio-visual tasks.

\section{Methods}
\label{sec:method}

\begin{figure}[tp]
    \centering
    \includegraphics[width=\linewidth]{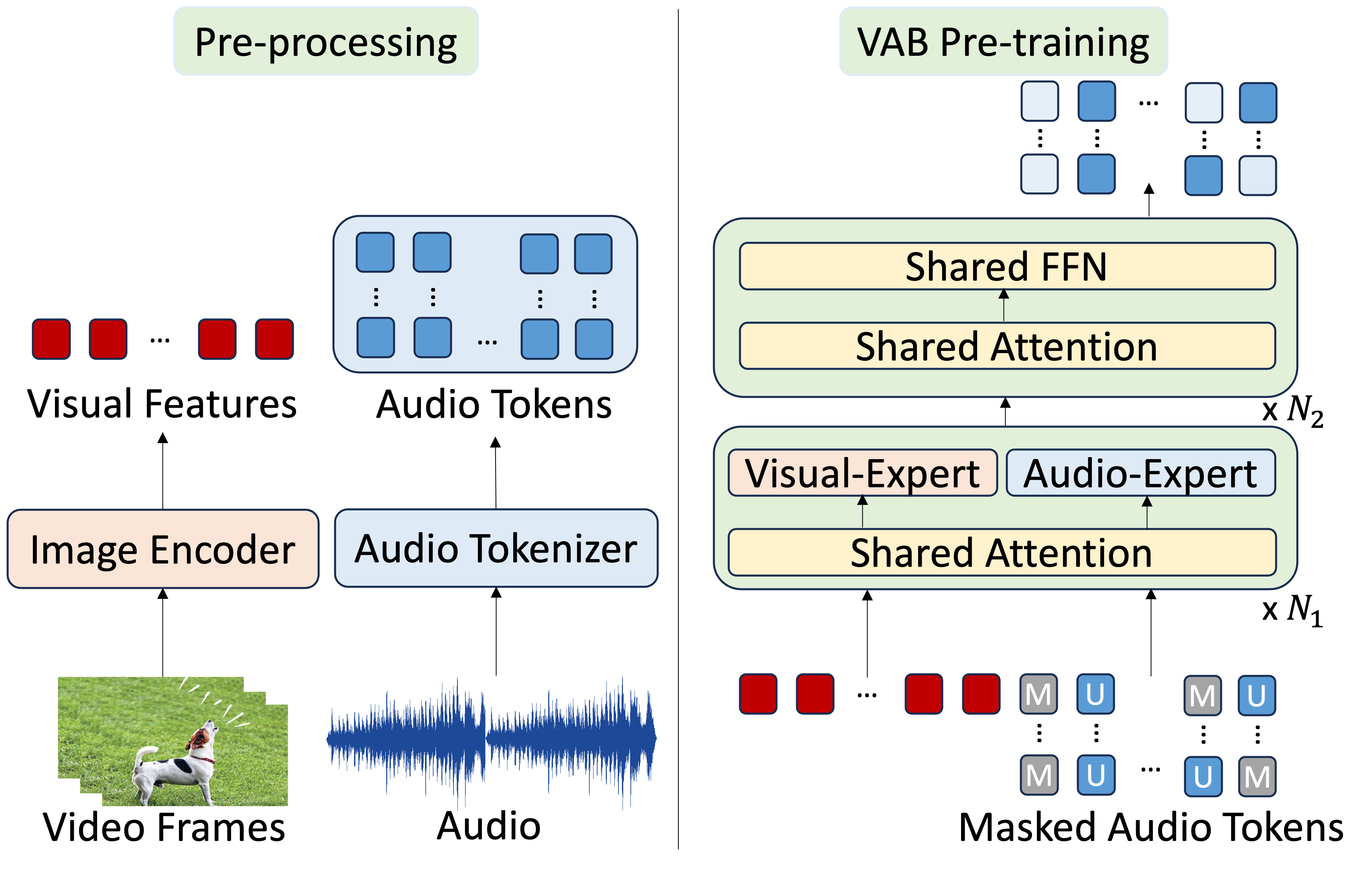}
    \vspace{-2mm}
    \caption{Pre-processing (left) and masked audio token prediction pre-training (right) of VAB framework.}
    \label{fig:vab_pretrain}
\end{figure}

In this section, we detail the design of Vision to Audio and Beyond (VAB) framework and its  capability of generating high-quality audio from silent video and acquiring semantic audio-visual representations for subsequent tasks. We first describe VAB \textit{pre-processing} stage that consists of converting raw audio and visual signals into latent spaces through the utilization of a pre-trained audio neural codec and image encoder. With audio tokens and frame-level visual features as inputs, we outline VAB components that facilitate \textit{self-supervised pre-training} stage centered around masked audio token prediction conditioned on visual features. At this stage VAB establishes its representation and learns to generate audio for video. In a subsequent stage, we elaborate on how the pre-trained VAB model is leveraged for \textit{vision-to-audio generation} and \textit{fine-tuning across various downstream tasks}.

\subsection{Audio and Video Transformation into Latent Spaces}
There are two primary motivations for converting audio and video frames into latent spaces. First, prior research~\cite{kreuk2022audiogen, copet2023simple, mei2023foleygen} indicates that performing generative modeling within latent spaces enhances training convergence and facilitates the generation of high-quality audio samples. Second, operating on raw video frames instead of latents,  entails significantly longer sequences and places a substantial computational burden. For instance, a ten-second video at 1fps results in 1960 patches when employing a standard Vision Transformer (ViT)~\cite{dosovitskiy2020image}. For such number of patches even employment of tubelets does not result in significant relief and therefore more compact representation through latents could be advantageous. We thus have utilized a frozen pre-trained image encoder, such as CLIP~\cite{radford2021learning}, to extract frame-level features and serve as latents. This choice enables a significant reduction in the length of visual sequences and benefits the semantic features obtained from image-text pre-training. In fact, we discovered that these well-established image-text features can be adapted to capture audio-video relationships effectively, thereby mitigating the potential information loss associated with compressed audio tokens.
Indeed, we use CLIP image encoder to extract frame-level features at a rate of 1fps, resulting in 10 seconds of video features $V \in \mathbb{R}^{10 \times d}$, where $d$ is the dimension of CLIP image features.
\\
To transform audio waveforms into tokens, we explore two off-the-shelf pre-trained neural codec variants, DAC~\cite{kumar2023high} and Encodec~\cite{defossez2022high}. DAC and Encodec are both trained in a fully self-supervised manner on reconstruction tasks without reliance on any labels and provide audio tokens that represent compressed versions of the original signals. Both DAC and Encodec use $K$ residual vector quantization (RVQ) to encode 1D audio waveform $A_w \in \mathbb{R}^{T_a}$ (16kHz) into audio tokens $A \in \mathbb{N}^{K \times S}$, where $K$ is the number of residual codebooks and $S = T_a / d_a$, $d_a=320$ is the downsample factor for both codec. A $10$ seconds audio results in $A \in \mathbb{N}^{K \times 500}$ tokens in total. In DAC, it comprises $K=12$ codebooks. These codebooks constitute a hierarchy wherein codes in lower levels represent coarse acoustic features, while codes in higher levels capture finer acoustic details. On the other hand, the Encodec contains only $K=4$ codebooks, exhibiting poorer audio reconstruction quality compared to the DAC (See Appendix~\ref{upperbound}). To conduct a comprehensive study, we explored the use of both DAC and Encodec tokens. During VAB pre-training, we employed the first four levels of audio tokens $A_c = A^{0:4, 500}$ for both DAC and Encodec. For the remaining $8$ levels of DAC tokens, we followed Vampnet~\cite{garcia2023vampnet}, applying an additional coarse-to-fine model solely for the purpose of audio generation.
\\
It is important to emphasize that both audio tokens and frame-level visual features are extracted before the VAB pre-training, which enables us to model temporal and audio-visual relationships during the pre-training process more efficiently.

\subsection{Conditional Masked Audio Token Prediction}
\textbf{Masking:} Given audio tokens $A_c$ and visual features $V$, we employ visual-conditioned masked audio tokens prediction as the VAB pre-training task. First, we randomly mask the audio tokens using a masking strategy with variable masking ratios. Specifically, we sample the masking ratio $M_r$ from a truncated Gaussian distribution centered at $0.55$ with $std = 0.25$, left truncated by $0.5$ and right truncated by $1$. It is important to have a variable and reasonable portion of masked audio tokens to enable learning both the representation and the generation. We reuse the codebook embeddings in neural codecs and add a new [\textit{MASK}] token. The masked audio tokens $A_m \in \mathbb{N}^{4, 500}$ are embedded and summed to obtain masked audio embeddings $A_{\text{emb}} \in \mathbb{R}^{500, d_{\text{emb}}}$, where $d_{\text{emb}}$ is the embedding dimension. The visual features $V$ are projected linearly to the same dimension as $d_{\text{emb}}$ to have visual embeddings $V_{\text{emb}}$ and to add modality-specific embeddings $E_v$ and $E_a$ to $V_{\text{emb}}$ and $A_{\text{emb}}$, respectively. We then concatenate the visual and audio embeddings to form the final input sequence $x = [V'_{\text{emb}}, A'_{\text{emb}}]$, where $V'_{\text{emb}} = V_{\text{emb}} + E_v$, $A'_{\text{emb}} = A_{\text{emb}} + E_a$.

\textbf{Multiway Transformer Encoder:} The architecture of VAB is similar to the Multiway Transformer Encoder~\cite{bao2022vlmo, wang2022image} where each transformer layer contains a bi-directional multi-head attention shared for audio and visual embeddings, layer normalization, feed-forward networks, and residual connections. For VAB model with $N$ layers, we use modal-specific feed-forward networks for the first $N_1$ layers and shared feed-forward networks for the rest of $N_2 = N - N_1$ layers. The shared bi-directional self-attention allows audio and visual embeddings to associate with each other during training. The modal-specific feed-forward networks can be considered experts in learning information specifically for audio or vision so that we can use the first $N_1$-th layers as either audio or visual encoders for single modality tasks. The upper-level joint feed-forward networks are useful for the more challenging vision-to-audio generation. Finally, multiple linear projection heads are used to predict each level of masked audio tokens.

Let $A_u$ be the set of all unmasked audio tokens and VAB model parameters $\theta$. The objective of the pre-training is to minimize the negative log-likelihood
\begin{equation}
    l_{\text{vab}} = - \sum_{\forall a \in A_m} \log p(a | A_u, V, \theta)
\end{equation}
The overview of VAB pre-processing and pre-training is shown in Fig~\ref{fig:vab_pretrain}.

\begin{figure*}[tp]
    \centering
    \includegraphics[width=\linewidth]{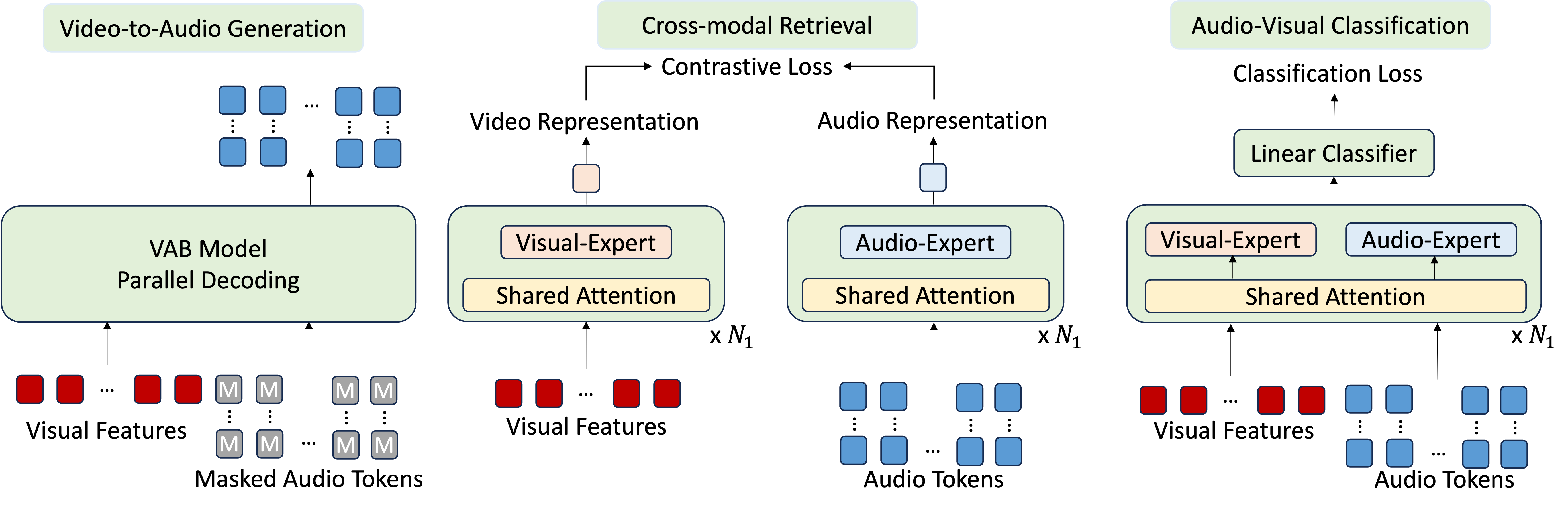}
    \vspace{-1mm}
    \caption{After the pre-training phase, VAB allows for zero-shot video-to-audio generation (left). Moreover, it can undergo representation adaptation fine-tuning to facilitate cross-modal retrieval through contrastive loss (middle) and accommodate classification tasks by incorporation of a linear classifier (right).}
    \label{fig:app}
\end{figure*}

\subsection{Zero-Shot Video-to-Audio Generation}
After VAB pre-training, audio tokens can be generated using efficient iterative decoding similar to previous masked generative modeling approaches~\cite{chang2022maskgit, li2023mage, garcia2023vampnet}. Specifically, we initialize all audio tokens by [\textit{MASK}] tokens and feed them into the VAB model along with the visual features $V$, similarly to pre-training. For each decoding iteration $t \in [0, t_T]$, the process involves predicting the token for the remaining masked audio token to obtain $\hat{a_t}$. Subsequently, we compute a confidence score $z$ by incorporating the prediction log-probabilities and introducing temperature-annealed Gumbel noise
\begin{equation}
    z(\hat{a_t}) = \log(p(\hat{a_t})) + \alpha \cdot g_t,
\end{equation}
where $g_t$ represents an independently and identically distributed (i.i.d) sample drawn from Gumbel noise, and $\alpha$ signifies the temperature that undergoes linear annealing, gradually reducing to $0$ as the number of sampling iterations progresses. Following this, we sort the set of sampled tokens based on their confidence scores and determine the $k$ lowest confidence tokens to re-mask in the subsequent sampling iteration. Specifically, the number of tokens to be re-masked, denoted as $k = \gamma (\frac{t}{t_T})N$, where $N$ represents the total number of tokens, and $\gamma$ follows a cosine schedule so that fewer masked audio tokens are replaced in the early iterations, while more masked audio tokens will be replaced in the later iterations. Additionally, classifier-free guidance~\cite{ho2022classifier} is also employed to achieve better visual adherence.

As a result, the sampled audio tokens are ready to be decoded by DAC or Encodec to generate audio waveforms. For DAC, the quality of the audio tokens decoded by coarse levels turns out to be insufficient~\cite{agostinelli2023musiclm, su2023v2meow, garcia2023vampnet}. Therefore, we train an additional coarse-to-fine model to generate fine-level audio tokens $A_f$ conditioned on coarse tokens $A_c$. The coarse-to-fine model architecture is a bi-directional transformer encoder as well. In comparison to the coarse model in the previous stage, we do not use modal-specific feed-forward networks, and we use all levels of audio tokens instead of coarse-level tokens as inputs. We train the model using masked audio token prediction, but only fine-level audio tokens would be masked and predicted. Formally, training the model $\theta_{\text{c2f}}$ minimizes the objective
\begin{equation}
    l_{\text{c2f}} = - \sum_{\forall a \in A_{f, m}} \log p(a | A_{f, u}, A_c, V, \theta_{\text{c2f}}).
\end{equation}
During inference, generated coarse audio tokens and visual features will be used as the conditions and iterative decoding will be used in coarse-to-fine again for generation. Unlike autoregressive generation, which generates tokens one by one, our approach generates them simultaneously. This significantly reduces the total number of inference steps while still achieving, and in some cases even surpassing, the performance of the autoregressive approach.

\subsection{Adaptation of VAB for Retrieval}
After VAB pre-training, the learned representation is amenable for adaptation to a variety of audio-visual tasks through fine-tuning. We therefore proceed to fine-tune the first $N_1$ modal-specific layers of the VAB model using the contrastive loss, aiming to align audio and visual modalities for retrieval tasks. While it might seem logical to apply masked prediction and contrastive loss concurrently during the VAB pre-training, our early experiments revealed that such a training strategy led to failure of both tasks and could not converge. Additionally, we observed that initializing contrastive training from scratch on the first $N_1$ layers required more training epochs to converge to a similar performance as the one initialized from pre-trained VAB with masked audio token prediction. Furthermore, fine-tuning for masked audio token prediction initiated from the contrastive pre-training model did not aid the convergence of masked audio tokens prediction (See Appendix~\ref{choice}). As a result of these insights, we apply contrastive training as fine-tuning task after the masked prediction pre-training phase. Specifically, we run two forward passes to obtain audio and visual output features from audio tokens and visual features, separately. The audio tokens are not masked at this stage. The resulting output features are then average-pooled and normalized. We fine-tune the model using the standard contrastive loss $L_c$
\begin{align}
    l_c = -\frac{1}{N}\sum^N_{i=1}\log \left[ \frac{\exp{s_{i,i}/ \tau}}{\sum_{k\neq i} \exp{s_{i,k}/ \tau} + \exp{s_{i,i}/ \tau}} \right],
\end{align}
where $s_{i,j} = || a^c_i ||^T || v^c_j ||$, $a^c$ and $v^c$ are audio and video representations, and $\tau$ is the learnable temperature value initialized with $0.05$. 

\subsection{Adaptation of VAB for Classification}
The latent representation can be adapted to additional tasks, for example, classification. For uni-modal classification tasks, we provide audio tokens or visual features as inputs to the first $N_1$ modal-specific layers of the VAB model. For audio-visual joint classification, we use both audio tokens and visual features as inputs and concatenate them into a joint sequence as in VAB pre-training. Similar to the contrastive fine-tuning, we do not mask audio tokens. We average-pool the output features of the $N_1$ layers and add a linear classifier for fine-tuning all classification tasks. While fine-tuned VAB pre-trained model achieves reasonable performances in the classification tasks, we found that the contrastive fine-tuned VAB model could serve as a better initialization for classification fine-tuning (Appendix~\ref{pretrain_effect}). This advantage may stem from the fact that both retrieval and classification tasks do not require masking of audio tokens, thereby enabling a more seamless transfer of knowledge. This effect is particularly evident in audio-only and audio-visual classification tasks.

\section{Experiments}
\label{sec:experiments}
% \subsection{Implementation}
\textbf{Implementation.} We conduct pre-training of VAB using a combination of AudioSet~\cite{gemmeke2017audio} and VGGSound~\cite{chen2020vggsound}, two common extensive audio-visual datasets sourced from YouTube. For AudioSet, we are able to gather about 1.57M videos from the original AudioSet-2M, since not all of them are available. In the case of VGGSound, we use the VGGSound training set comprising about 177K videos. To extract audio tokens, we use off-the-shelf pre-trained Encodec and DAC models, processing audio at 16kHz sampling rate. For video frames, we adopt the pre-trained eva-CLIP image encoder, as employed in BLIP~\cite{li2023blip, fang2023eva}, to extract CLIP embeddings at a rate of 1fps. Both audio tokens and CLIP embeddings were pre-processed and stored prior to training the VAB model. Our experimentation encompassed two model sizes: VAB-Encodec/VAB-DAC, employed for reporting and comparison across all tasks, and VAB-DAC-Test, used for ablation studies and in-depth analysis. Both VAB-Encodec and VAB-DAC models comprise a total of 24 layers, with each layer consisting of 1024 dimensions and 16 attention heads (403M params). The first 12 layers are dedicated to modal-specific experts (253M params).

For the VAB-DAC-Test, all transformer layers are 768-dimensional with 12 attention heads, totaling 20 layers, where the first 12 layers contain modal-specific experts. For pre-training of VAB, we use AdamW optimizer with a target learning rate of 2e-4 and a cosine scheduler. All VAB pre-training, as well as the fine-tuning for contrastive loss, were conducted using 1 A100 (80G) GPU. For subsequent classification tasks, we used 1 A100 (40G) GPU. For additional details of model architecture and training configurations, please refer to the Appendix~\ref{model_arch} and~\ref{train_config}.

\begin{table}[t]
\footnotesize
\centering
\begin{tabular}{lcccccc}
\toprule
Methods & FAD $\downarrow$ & KLD $\downarrow$ & Speed (s) $\downarrow$\\
\midrule
SpecVQGAN & 6.63 & 3.78  & 7.2 \\
IM2WAV       & 6.32 & \bf{2.54}  & 289.5 \\
Diff-Foley & 6.40 & 3.15  & 4.4 \\
FoleyGen & \bf{2.59} & 2.89  & 6.9 \\
\cellcolor{mygray-bg}\bf{VAB-DAC (Ours)} & \cellcolor{mygray-bg} 3.24 & \cellcolor{mygray-bg} 2.84  & \cellcolor{mygray-bg} 1.3 \\
\cellcolor{mygray-bg}\bf{VAB-Encodec (Ours)} & \cellcolor{mygray-bg} 2.69 & \cellcolor{mygray-bg} 2.58 & \cellcolor{mygray-bg}\bf{0.4} \\

\bottomrule
\end{tabular}
\caption{Quantitative evaluation for video-to-audio generation on the VGGSound test set. Values in bold indicate the best value.}
\label{tab:gen}
\end{table}
\begin{figure}[t]
    \centering
    \includegraphics[width=\linewidth]{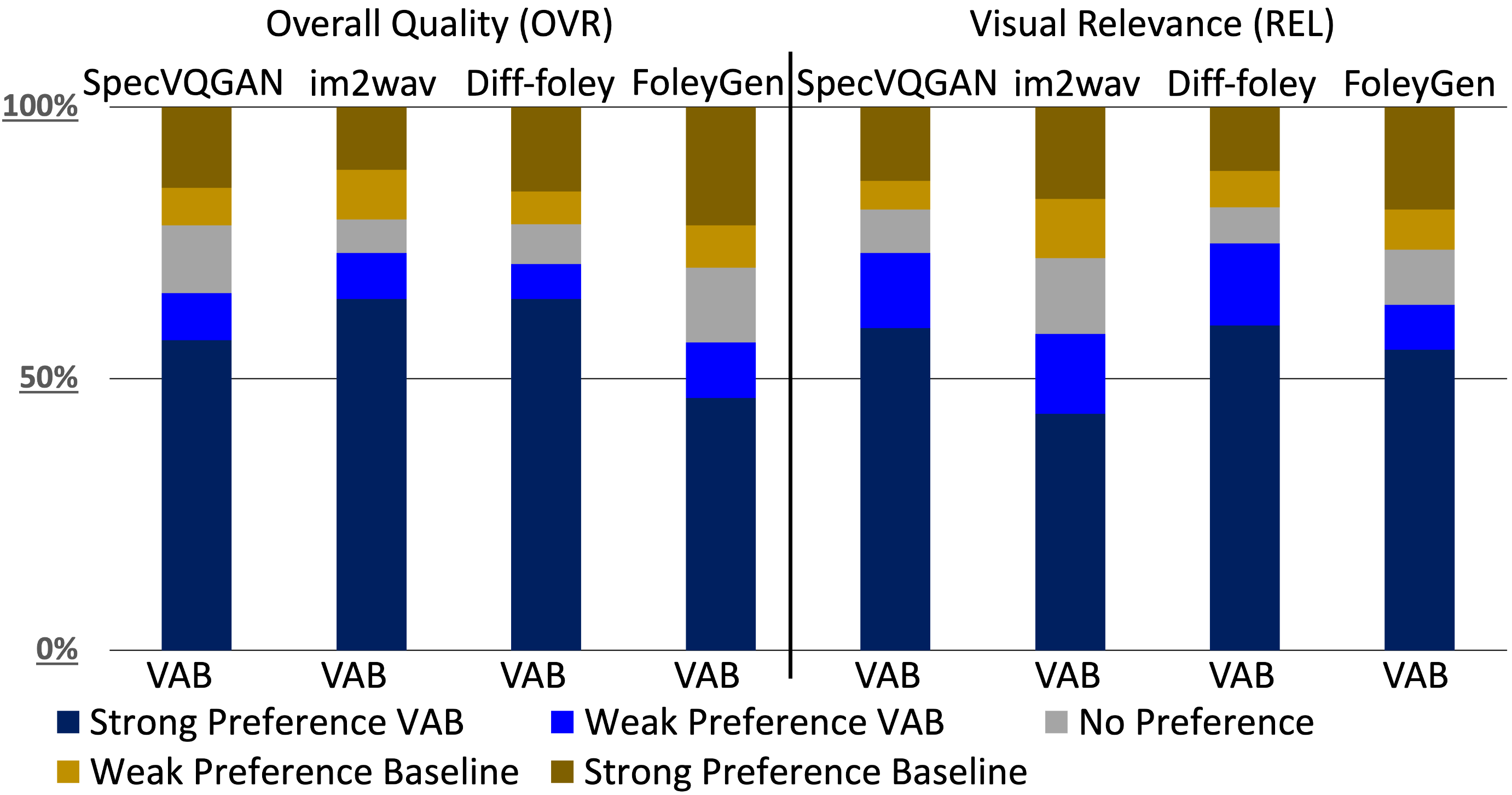}
    \vspace{-2mm}
    \caption{Human evaluations for video-to-audio generation. Left: overall quality of the video. Right: visual relevance to the audio.}
    \label{fig:human_eval}
\end{figure}

\begin{table}[t]
\footnotesize
\centering
\begin{tabular}{lcc}
\toprule
Methods & MOS mean $\uparrow$ & MOS std\\
\midrule
SpecVQGAN & 2.48 & 1.35\\
Img2Wav & 2.67 & 1.34\\
Diff-Foley & 2.62 & 1.35\\
FoleyGen & 2.76 & 1.35\\
\cellcolor{mygray-bg}\bf{VAB-Encodec (Ours)} & \cellcolor{mygray-bg} \bf{2.86} & \cellcolor{mygray-bg} 1.40\\
\bottomrule
\end{tabular}
\caption{Mean Opinion Score (MOS) for video-to-audio generation. Values in bold indicate the best value.}
\label{table:mos}
\end{table}
\textbf{Zero-Shot Video-to-Audio Generation.}
We evaluated video-to-audio generation using VGGSound test set, comparing our models against existing baselines: SpecVQGAN~\cite{iashin2021taming}, IM2WAV~\cite{sheffer2023hear}, Diff-Foley~\cite{luo2023diff}, and FoleyGen~\cite{mei2023foleygen}. We generate 10 seconds of audio for the entire VGGSound test set, resulting in 15546 samples. For VAB-Encodec and VAB-DAC, we utilize 16 decoding steps and set the classifier-free guidance scale to 5. In the case of the DAC coarse-to-fine model, we employed 36 decoding steps. To quantitatively assess the audio generation quality, we use two objective metrics: FAD (Fréchet Audio Distance)~\cite{kilgour2018fr} and KLD (Kullback–Leibler Distance). FAD is a collection-based metric that measures the similarity between the generated audio features and ground truth audio features extracted by the VGGish Network~\cite{hershey2017cnn} trained on AudioSet. In contrast, KLD quantifies the individual differences between each generated audio sample and ground truth, based on predicted label distribution extracted from a pre-trained PaSST model~\cite{koutini2021efficient}. FAD exhibits a correlation with human perception of audio quality, while KLD reflects the underlying audio categories conveyed in the sample. Additionally, we compared the inference speed of each video-to-audio model. The reported metric is the average time required to generate one 10-second audio sample from a video using one A6000 GPU. For SpecVQGAN, IM2WAV, and Diff-Foley, we adopted their open-sourced pre-trained models to generate samples. For FoleyGen, we closely followed the publication to reproduce the model. For further details regarding baseline sample generation and FoleyGen implementation, please refer to the Appendix~\ref{baseline_gen}. 
The quantitative results are presented in the Table~\ref{tab:gen}. A key observation is that modeling using state-of-the-art audio discrete tokens with multi-codebooks (FoleyGen and VAB) yields superior audio quality than previous methods. With the masked audio tokens prediction task and iterative-decoding approach, VAB achieved significantly, of order of magnitude, faster inference speeds compared to all previous works. While both FAD and KLD scores are both slightly lower than the best approaches, VAB achieves a more balanced performance in both metrics.

To gain a deeper insight into the perceptual differences between VAB generated samples and those of the baselines, we conducted subjective evaluations through two surveys. The first survey primarily assessed two subjective aspects of generated audio: OVR (overall quality) and REL (visual relevance). To ensure the robustness of our findings, we randomly selected 150 samples from a class-balanced subset of the VGGSound test set for each method. The study utilizes an A-vs-B human rating task, where raters are presented with two samples: one generated by one of the baseline models and the other by best VAB-Encodec model. Raters provide one of five possible responses, indicating a strong or weak preference for either A or B, or expressing no preference. Details of the subjective study are included in appendix~\ref{human_eval}. The results are shown in Fig.~\ref{fig:human_eval}. Notably, VAB-Encodec model was consistently more preferred over-all, in comparison to other methods in both questions, even those that scored better on FAD or KLD metrics. In the second survey, we conducted a Mean Opinion Score (MOS) comparison across each model. Utilizing the same generated samples as in the first survey. The results, including mean and standard deviation, are presented in the Table~\ref{table:mos}. These MOS results further highlight the advantages of the VAB model.

\begin{table*}[th]
\footnotesize
\centering
\begin{tabular}{lccccccccc}
\toprule
&\multicolumn{3}{c}{AudioSet Eval Subset} & \multicolumn{3}{c}{VGGSound Eval Subset} & \multicolumn{3}{c}{MSR-VTT (Zero-shot)} \\
& R@1 & R@5 & R@10 & R@1 & R@5 & R@10 & R@1 & R@5 & R@10 \\
\midrule
\underline{\textit{Video $\rightarrow$ Audio}} & & & & & & & & & \\
LanguageBind~\cite{zhu2023languagebind} & 6.4 & 20.2 & 28.3 & 10.3 & 30.1 & 39.7 & 1.9 & 6.2 & 8.8\\
ImageBind~\cite{girdhar2023imagebind} & 22.1 & 43.2 & 52.6 & 21.6 & 43.4 & 52.9 & 7.0 & 18.5 & 25.2\\
CAV-MAE~\cite{gong2022contrastive} & 18.8 & 39.5  & 50.1  & 14.8 & 34.2  & 44.0 & 13.3 & 29.0 & 40.5 \\
\cellcolor{mygray-bg}\bf{VAB-DAC (Ours)} & \cellcolor{mygray-bg}35.5 & \cellcolor{mygray-bg}61.8 & \cellcolor{mygray-bg} 72.4 & \cellcolor{mygray-bg} 30.8 & \cellcolor{mygray-bg} 59.6 & \cellcolor{mygray-bg} 70.8 & \cellcolor{mygray-bg} \bf{13.8}  & \cellcolor{mygray-bg} 30.6 & \cellcolor{mygray-bg} 40.1\\
\cellcolor{mygray-bg}\bf{VAB-Encodec (Ours)} & \cellcolor{mygray-bg}\bf{39.5} & \cellcolor{mygray-bg}\bf{65.4} & \cellcolor{mygray-bg}\bf{74.6} & \cellcolor{mygray-bg}\bf{33.5} & \cellcolor{mygray-bg}\bf{63.3} & \cellcolor{mygray-bg}\bf{74.3} & \cellcolor{mygray-bg} 14.2 & \cellcolor{mygray-bg}\bf{31.1} & \cellcolor{mygray-bg}\bf{42.0}\\
\midrule
\underline{\textit{Audio $\rightarrow$ Video}} & & & & & & & & & \\
LanguageBind~\cite{zhu2023languagebind} & 4.4 & 15.0 & 22.5 & 6.5 & 22.7 & 33.5 & 1.2 & 4.4 & 6.9\\
ImageBind~\cite{girdhar2023imagebind}& 20.8 & 42.6 & 51.6 & 20.7 & 43.2 & 53.4 & 6.0 & 16.9 & 23.7\\
CAV-MAE~\cite{gong2022contrastive} & 15.1 & 34.0 & 43.0 & 12.8 & 30.4 & 40.3 & 7.6 & 19.8 & 30.2 \\
\cellcolor{mygray-bg}\bf{VAB-DAC (Ours)} & \cellcolor{mygray-bg} 37.0 & \cellcolor{mygray-bg} 61.8 & \cellcolor{mygray-bg} 70.8 & \cellcolor{mygray-bg} 33.1 & \cellcolor{mygray-bg}\bf{62.7} & \cellcolor{mygray-bg}\bf{73.8} & \cellcolor{mygray-bg} \bf{12.0} & \cellcolor{mygray-bg}\bf{27.3} & \cellcolor{mygray-bg}\bf{36.2}\\
\cellcolor{mygray-bg}\bf{VAB-Encodec (Ours)} & \cellcolor{mygray-bg} \bf{37.5} & \cellcolor{mygray-bg}\bf{64.0} & \cellcolor{mygray-bg}\bf{73.7} & \cellcolor{mygray-bg}\bf{34.9} & \cellcolor{mygray-bg} \bf{62.7} & \cellcolor{mygray-bg}\bf{73.1} & \cellcolor{mygray-bg} 9.6 & \cellcolor{mygray-bg} 23.3 & \cellcolor{mygray-bg} 32.9\\
\bottomrule
\end{tabular}
\caption{Cross-modal retrieval results on AudioSet, VGGSound, and MSR-VTT. Values in bold highlight the best performance.}
\label{tab:retrieval}
\end{table*}
\textbf{Audio-Visual Retrieval.}
In this task, we assess the learned representations of VAB model for both audio-to-visual retrieval and visual-to-audio retrieval. We first fine-tune the VAB model using the contrastive loss, employing the same training dataset utilized during the pre-training phase. To evaluate the retrieval performance, we adopt the CAV-MAE~\cite{gong2022contrastive} methodology for conducting retrieval on audio-visual samples sourced from the AudioSet and VGGSound evaluation set. Furthermore, we extend our evaluation to include zero-shot retrieval on MSR-VTT~\cite{xu2016msr} test set. We feed audio tokens and visual features through the VAB model in two separate forward passes. Subsequently, we compute the mean-pooled and normalized encoder outputs to derive audio and visual representations, respectively. We then calculate retrieval recall metrics at ranks 1, 5, and 10 (R@1, R@5, R@10) based on the cosine similarity of these audio and visual representations. Besides CAV-MAE, we expanded our comparison of retreival performances by comparing VAB with two additional open-source multi-modal alignment models: ImageBind~\cite{girdhar2023imagebind} and LanguageBind~\cite{zhu2023languagebind}. These models primarily utilize images or text, respectively, as the main modality to unify various modalities within the latent space. The comparison results are presented in the Table~\ref{tab:retrieval}. Notably, we found that VAB models performed consistently better on Audioset and VGGSound than all of the baselines by a large margin (x2 improvement on AudioSet and VGGSound). We also discovered that the video-to-audio retrieval is generally better than the audio-to-video retrieval for all datasets. We further observed that while LanguageBind is known for its effective performance in text-centric tasks compared to ImageBind, its effectiveness does not generalize well in correlating audio and visual modalities.

\textbf{Audio-Visual Event Classification.}
\begin{table*}[th]
\centering
\footnotesize
\begin{tabular}{lccccccccccc}
\toprule
& \multicolumn{3}{c}{\underline{VGGSound (Acc)}$\uparrow$} & \multicolumn{3}{c}{\underline{AS-2M (mAP)}$\uparrow$} &\multicolumn{3}{c}{\underline{AS-20K (mAP)}$\uparrow$}  \\
Method & V+A & V & A & V+A & V & A & V+A & V & A \\
\midrule

\multicolumn{3}{l}{\underline{\textit{Audio-visual Models}}} & & & & & & \\
G-Blend~\cite{wang2020makes} & - & - & - & 41.8 & 18.8 & 32.4 & 37.8 & 22.1 & 29.1 \\
Perceiver~\cite{jaegle2021perceiver} & - & - & - & 44.2 & 25.8 & 38.4 & - & - & - \\
Attn AV~\cite{fayek2020large} & - & - & - & 44.2 & 25.7 & 38.4 & - & - & - \\
CAV-MAE~\cite{gong2022contrastive} & 65.5 & 47.0 & 59.5 & 51.2 & 26.2 & 46.6 & 42.0 & 19.8 & 37.7 \\
MBT~\cite{nagrani2021attention} & 64.1 & 51.2 & 52.3 & 49.6 & 31.3 & 41.5 & 43.9 & 27.7 & 31.3 \\
% MAViL & 41.6 & 23.7 & 44.6 & 48.7 & 28.3 & 51.9 & 60.6 & 50.0 & 66.5 \\
MAViL~\cite{huang2022mavil} & \bf{67.1} & 50.9 & \bf{60.8} & \bf{53.3} & 30.3 & \bf{48.7} & \bf{44.9} & 24.8 & \bf{41.8} \\
\midrule
\cellcolor{mygray-bg}\bf{VAB-DAC (Ours)} & \cellcolor{mygray-bg} 63.9 & \cellcolor{mygray-bg}\bf{55.4} & \cellcolor{mygray-bg}48.2 & \cellcolor{mygray-bg} 47.0 & \cellcolor{mygray-bg} 33.3 & \cellcolor{mygray-bg} 36.2 & \cellcolor{mygray-bg} 38.9 & \cellcolor{mygray-bg} 28.3 & \cellcolor{mygray-bg} 28.8\\
\cellcolor{mygray-bg}\bf{VAB-Encodec (Ours)} & \cellcolor{mygray-bg} 65.2 & \cellcolor{mygray-bg} 55.1 & \cellcolor{mygray-bg} 51.3 & \cellcolor{mygray-bg} 47.7 & \cellcolor{mygray-bg}\bf{33.5} & \cellcolor{mygray-bg} 38.6 & \cellcolor{mygray-bg} 38.7 & \cellcolor{mygray-bg}\bf{29.0} & \cellcolor{mygray-bg} 29.0\\
\bottomrule
\end{tabular}
\caption{Comparison to previous audio-visual models on VGGSound, AS-2M, AS-20K in audio-visual (V+A), video-only (V) and audio-only (A) classification tasks. Values in bold represent the best performance.}
\label{tab:av_cls_results}
\end{table*}
In this task, we evaluate the quality of VAB representations in the context of audio-visual event classification task. To accomplish this, we employ the contrastive VAB model and fine-tune it on three distinct datasets: 1) AudioSet-20K, 2) AudioSet-2M, and 3) VGGSound. During the classification fine-tuning stage, we retain the first $N_1$ layers of the model and add a linear classification head. Our model is fine-tuned using audio-only data (A), video-only data (V), and audio-visual data (V+A), enabling us to evaluate both single-modal and multi-modal representation quality. The results of the evaluation are presented in the Table~\ref{tab:av_cls_results}. Both VAB-DAC and VAB-Encodec exhibit competitive performances across A, V, and V+A classification tasks, with a slight advantage for VAB-Encodec. Notably, visual-only (V) classification outperforms previous methods across three datasets, highlighting the effectiveness of incorporating frame-level CLIP embeddings as our visual features. Additionally, we observed a notable performance gap in the audio-only (A) when compared to the best-performing methods. This outcome can be attributed to the fact that audio tokens represent a lossy compression of the original audio. Without the guidance of visual features, correctly categorizing audio becomes a more challenging task. This observation aligns with findings in self-supervised learning in the image domain using quantized tokens~\cite{li2023mage}. However, as demonstrated in Appendix~\ref{from_scratch}, VAB pre-training significantly improves supervised training using visual features and audio tokens from scratch.

\textbf{Audio-only Classification.}
To assess the generalization of the acquired audio representations, we further evaluate the pre-trained VAB model by transferring it to other speech-only or audio-only tasks outside its original domain. In particular, we follow MAViL~\cite{huang2022mavil} and conduct experiments on the Environmental Sound Classification (ESC-50)~\cite{piczak2015esc} and Speech Commands (SPC-v1)~\cite{warden2018speech} datasets. In these experiments, only the audio branch of VAB is fine-tuned. The results, presented in Table~\ref{tab:audio_cls}, demonstrate that VAB achieves competitive performance to recent supervised and self-supervised models. These findings underscore the adaptability and transferability of VAB, as it can seamlessly transition from audio-visual self-supervised pre-training to audio-only downstream tasks.

\begin{table}[ht]
\footnotesize
\centering
\begin{tabular}{lcc}
\toprule
Method & ESC-50 & SPC-1 \\
\midrule
AST~\cite{gong2021ast} & 88.7 & 95.5 \\
SS-AST~\cite{gong2022ssast} & 88.8 & 96.0 \\
Aud-MAE~\cite{huang2022masked} & 94.1 & 96.9 \\
MAViL~\cite{huang2022mavil} & \bf{94.4} & \bf{97.4} \\
\cellcolor{mygray-bg}\bf{VAB-DAC (Ours)} & \cellcolor{mygray-bg} 89.2 & \cellcolor{mygray-bg} 95.1 \\
\cellcolor{mygray-bg}\bf{VAB-Encodec (Ours)} & \cellcolor{mygray-bg} 91.4 & \cellcolor{mygray-bg} 96.1 \\
\bottomrule
\end{tabular}
\caption{Comparison with ESC-50 and SPC-1 audio only classification accuracy.}
\label{tab:audio_cls}
\end{table}

\begin{table*}[!thb]
\centering
\begin{tabular}{lccccccc}

\toprule
& \multicolumn{3}{c}{\underline{VGGSound (Acc)}$\uparrow$} & \multicolumn{3}{c}{\underline{AS-20K (mAP)}$\uparrow$}  \\
Method & V+A & V & A & V+A & V & A \\
\midrule
VAB-Encodec Linear Probe & 57.6 & 53.4 & 48.7 & 33.3 & 26.9 & 28.9 \\
CAV-MAE Linear Probe & 54.2 & - & - & 29.8 & - & - \\
 \cellcolor{mygray-bg} VAB-Encodec Fine-tuning &  \cellcolor{mygray-bg} \bf{65.2} & \cellcolor{mygray-bg} \bf{55.1} & \cellcolor{mygray-bg} \bf{51.3} & \cellcolor{mygray-bg} \bf{38.7} & \cellcolor{mygray-bg} \bf{29.0} & \cellcolor{mygray-bg} \bf{29.0} \\ \hline
\end{tabular}
\caption{Linear Probing comparison results.}
\label{tab:linear_probe}
\end{table*}

\textbf{Linear Probing} To evaluate how much effort VAB model requires to adapt to downstream tasks, we include linear probing comparisons against full fine-tuned VAB model and CAV-MAE model with linear probing on VGGSound and AS-20K classification benchmark, detailed in Table \ref{tab:linear_probe}. The findings suggest that while linear probing exhibits lower performance compared to end-to-end fine-tuning, the discrepancy is within a reasonable margin. Additionally, we observed that the reduction in performance for the V+A (audio-visual) classification task is slightly larger than that for a single modality. This suggests that multimodal tasks may demand more extensive adaptation efforts.

\noindent\textbf{Analysis.}
We validated VAB structure and setup by performing comprehensive studies and ablations. We include the key outcomes of the studies in this section, and refer to the details in the appendix due to the page limit. 

Since contrastive learning and masking predictions are both self-supervised approaches,
we investigated the scenario of applying contrastive learning as the pre-training task. When compared with the model initialized by masked token prediction in pre-training, contrastive training turns out to take more time to converge to the same level of retrieval performance. The experimental results and details are presented in Appendix~\ref{choice}. Furthermore, we performed ablation studies for VAB architecture, specifically whether the modal-specific experts are critical (Appendix~\ref{expert}). Indeed, we find that using non-experts in the pre-training, significantly degrades the performance on classification tasks. We also investigated how masking ratio, used in pre-training task, affects the performance of downstream tasks (Appendix~\ref{mask_strategy}) and studied the effect of using different visual encoders and how it affects the VAB model performance (Appendix~\ref{visual_encoders}). Furthermore, we examined various configurations of video-to-audio generation during inference such as classifier-free guidance scale, masking temperature, and number of decoding steps (Appendix~\ref{gen_analysis}).

\section{Conclusion}
\label{sec:conclusion}
In this work, we present VAB, a general-purpose audio-visual framework that connects audio-visual representation learning and vision-to-audio generation, two tasks in audio-visual learning that have been disconnected thus far. VAB employs a pre-training of visual-conditioned masked audio token prediction, and obtains a uniform visual-audio model. Such an approach fosters both contextual learning and also empowers fast and high-quality video-to-audio generation. VAB can be adapted and lead to competitive performance across various audio-visual downstream tasks, including audio-visual event classification, audio-visual retrieval, and audio-only classification. 

\section*{Acknowledgements} We acknowledge the support of HDR Institute: Accelerated AI Algorithms for Data-Driven Discovery (A3D3) National Science Foundation grant PHY-2117997.

\section*{Impact Statement}
Conditional generative and unified models, such as VAB, have the potential to serve as a foundation for innovative tools, technologies, and practices that empower content creators. While our primary motivation is to aid creators in enhancing their creative endeavors, we recognize the imperative need for developing and deploying these models in a manner that diligently considers the values and well-being of creators, their communities, and society at large.

Notably, generative models inherently learn to replicate patterns and biases within their training data. Specifically, since VAB is as a unified and foundation model, it could have the capacity to perpetuate potential biases present in the video and audio used for its training. These biases can be subtle and challenging to detect, often eluding comprehensive assessment by current evaluation benchmarks. As a consequence, model-generated content may inadvertently express through audio or video demeaning or offensive language, stemming from learned associations or chance occurrences. 

Upon conducting a thorough analysis of the training dataset, we observed a skewed distribution of audio-visual events towards a few specific categories. Furthermore, within each genre, gender, age, or ethical groups were not consistently represented. For instance, male representation dominates certain music genres such as hip-hop and heavy metal. These concerns extend to learned visual-audio associations which may foster stereotypical links between video content (such as individuals, body movements, locations, and objects) and a limited set of audio events. Addressing these issues necessitates fairness testing to gauge the likelihood of such patterns in a given model and to intervene effectively.

In parallel with algorithmic advancements, we are actively engaged in initiatives aimed at the comprehension and mitigation of the potential risks of bias inherited from training data, issues related to cultural appropriation, and stereotyping. Further efforts are required to assess whether the audio generated is contextually appropriate, a determination that transcends technical assessments such as audio quality or visual relevance. This endeavor mandates a profound understanding of the social and audio context and is best undertaken in collaboration with cultural and audio experts. We emphasize that these concerns, among others, hold important value, on par with the algorithmic advancements that sometimes take precedence. It is imperative to consider the broader context in which these models operate.

\bibliography{references}

\begin{thebibliography}{72}
\providecommand{\natexlab}[1]{#1}
\providecommand{\url}[1]{\texttt{#1}}
\expandafter\ifx\csname urlstyle\endcsname\relax
  \providecommand{\doi}[1]{doi: #1}\else
  \providecommand{\doi}{doi: \begingroup \urlstyle{rm}\Url}\fi

\bibitem[Agostinelli et~al.(2023)Agostinelli, Denk, Borsos, Engel, Verzetti, Caillon, Huang, Jansen, Roberts, Tagliasacchi, Sharifi, Zeghidour, and Frank]{agostinelli2023musiclm}
Agostinelli, A., Denk, T.~I., Borsos, Z., Engel, J., Verzetti, M., Caillon, A., Huang, Q., Jansen, A., Roberts, A., Tagliasacchi, M., Sharifi, M., Zeghidour, N., and Frank, C.
\newblock Musiclm: Generating music from text, 2023.

\bibitem[Alayrac et~al.(2022)Alayrac, Donahue, Luc, Miech, Barr, Hasson, Lenc, Mensch, Millican, Reynolds, et~al.]{alayrac2022flamingo}
Alayrac, J.-B., Donahue, J., Luc, P., Miech, A., Barr, I., Hasson, Y., Lenc, K., Mensch, A., Millican, K., Reynolds, M., et~al.
\newblock Flamingo: a visual language model for few-shot learning.
\newblock \emph{Advances in Neural Information Processing Systems}, 35:\penalty0 23716--23736, 2022.

\bibitem[Arandjelovic \& Zisserman(2017)Arandjelovic and Zisserman]{arandjelovic2017look}
Arandjelovic, R. and Zisserman, A.
\newblock Look, listen and learn.
\newblock In \emph{Proceedings of the IEEE international conference on computer vision}, pp.\  609--617, 2017.

\bibitem[Arandjelovic \& Zisserman(2018)Arandjelovic and Zisserman]{arandjelovic2018objects}
Arandjelovic, R. and Zisserman, A.
\newblock Objects that sound.
\newblock In \emph{Proceedings of the European conference on computer vision (ECCV)}, pp.\  435--451, 2018.

\bibitem[Aytar et~al.(2016)Aytar, Vondrick, and Torralba]{aytar2016soundnet}
Aytar, Y., Vondrick, C., and Torralba, A.
\newblock Soundnet: Learning sound representations from unlabeled video.
\newblock \emph{Advances in neural information processing systems}, 29, 2016.

\bibitem[Bao et~al.(2022)Bao, Wang, Dong, Liu, Mohammed, Aggarwal, Som, Piao, and Wei]{bao2022vlmo}
Bao, H., Wang, W., Dong, L., Liu, Q., Mohammed, O.~K., Aggarwal, K., Som, S., Piao, S., and Wei, F.
\newblock Vlmo: Unified vision-language pre-training with mixture-of-modality-experts.
\newblock \emph{Advances in Neural Information Processing Systems}, 35:\penalty0 32897--32912, 2022.

\bibitem[Berglund et~al.(2015)Berglund, Raiko, Honkala, K\"{a}rkk\"{a}inen, Vetek, and Karhunen]{NIPS2015_c75b6f11}
Berglund, M., Raiko, T., Honkala, M., K\"{a}rkk\"{a}inen, L., Vetek, A., and Karhunen, J.~T.
\newblock Bidirectional recurrent neural networks as generative models.
\newblock In Cortes, C., Lawrence, N., Lee, D., Sugiyama, M., and Garnett, R. (eds.), \emph{Advances in Neural Information Processing Systems}, volume~28. Curran Associates, Inc., 2015.
\newblock URL \url{https://proceedings.neurips.cc/paper_files/paper/2015/file/c75b6f114c23a4d7ea11331e7c00e73c-Paper.pdf}.

\bibitem[Borsos et~al.(2023)Borsos, Sharifi, Vincent, Kharitonov, Zeghidour, and Tagliasacchi]{borsos2023soundstorm}
Borsos, Z., Sharifi, M., Vincent, D., Kharitonov, E., Zeghidour, N., and Tagliasacchi, M.
\newblock Soundstorm: Efficient parallel audio generation.
\newblock \emph{arXiv preprint arXiv:2305.09636}, 2023.

\bibitem[Chang et~al.(2022)Chang, Zhang, Jiang, Liu, and Freeman]{chang2022maskgit}
Chang, H., Zhang, H., Jiang, L., Liu, C., and Freeman, W.~T.
\newblock Maskgit: Masked generative image transformer.
\newblock In \emph{Proceedings of the IEEE/CVF Conference on Computer Vision and Pattern Recognition}, pp.\  11315--11325, 2022.

\bibitem[Chang et~al.(2023)Chang, Zhang, Barber, Maschinot, Lezama, Jiang, Yang, Murphy, Freeman, Rubinstein, et~al.]{chang2023muse}
Chang, H., Zhang, H., Barber, J., Maschinot, A., Lezama, J., Jiang, L., Yang, M.-H., Murphy, K., Freeman, W.~T., Rubinstein, M., et~al.
\newblock Muse: Text-to-image generation via masked generative transformers.
\newblock \emph{arXiv preprint arXiv:2301.00704}, 2023.

\bibitem[Chen et~al.(2020)Chen, Xie, Vedaldi, and Zisserman]{chen2020vggsound}
Chen, H., Xie, W., Vedaldi, A., and Zisserman, A.
\newblock Vggsound: A large-scale audio-visual dataset, 2020.

\bibitem[Copet et~al.(2023)Copet, Kreuk, Gat, Remez, Kant, Synnaeve, Adi, and D{\'e}fossez]{copet2023simple}
Copet, J., Kreuk, F., Gat, I., Remez, T., Kant, D., Synnaeve, G., Adi, Y., and D{\'e}fossez, A.
\newblock Simple and controllable music generation.
\newblock \emph{arXiv preprint arXiv:2306.05284}, 2023.

\bibitem[D{\'e}fossez et~al.(2022)D{\'e}fossez, Copet, Synnaeve, and Adi]{defossez2022high}
D{\'e}fossez, A., Copet, J., Synnaeve, G., and Adi, Y.
\newblock High fidelity neural audio compression.
\newblock \emph{arXiv preprint arXiv:2210.13438}, 2022.

\bibitem[Devlin et~al.(2019)Devlin, Chang, Lee, and Toutanova]{devlin2019bert}
Devlin, J., Chang, M.-W., Lee, K., and Toutanova, K.
\newblock Bert: Pre-training of deep bidirectional transformers for language understanding, 2019.

\bibitem[Dhariwal et~al.(2020)Dhariwal, Jun, Payne, Kim, Radford, and Sutskever]{dhariwal2020jukebox}
Dhariwal, P., Jun, H., Payne, C., Kim, J.~W., Radford, A., and Sutskever, I.
\newblock Jukebox: A generative model for music.
\newblock \emph{arXiv preprint arXiv:2005.00341}, 2020.

\bibitem[Di et~al.(2021)Di, Jiang, Liu, Wang, Zhu, He, Liu, and Yan]{di2021video}
Di, S., Jiang, Z., Liu, S., Wang, Z., Zhu, L., He, Z., Liu, H., and Yan, S.
\newblock Video background music generation with controllable music transformer.
\newblock In \emph{Proceedings of the 29th ACM International Conference on Multimedia}, pp.\  2037--2045, 2021.

\bibitem[Dong et~al.(2024)Dong, Liu, Chen, Polak, and Zhang]{Dong_2024_CVPR}
Dong, Z., Liu, X., Chen, B., Polak, P., and Zhang, P.
\newblock Musechat: A conversational music recommendation system for videos.
\newblock In \emph{Proceedings of the IEEE/CVF Conference on Computer Vision and Pattern Recognition (CVPR)}, pp.\  12775--12785, June 2024.

\bibitem[Dosovitskiy et~al.(2020)Dosovitskiy, Beyer, Kolesnikov, Weissenborn, Zhai, Unterthiner, Dehghani, Minderer, Heigold, Gelly, et~al.]{dosovitskiy2020image}
Dosovitskiy, A., Beyer, L., Kolesnikov, A., Weissenborn, D., Zhai, X., Unterthiner, T., Dehghani, M., Minderer, M., Heigold, G., Gelly, S., et~al.
\newblock An image is worth 16x16 words: Transformers for image recognition at scale.
\newblock \emph{arXiv preprint arXiv:2010.11929}, 2020.

\bibitem[Fang et~al.(2023)Fang, Sun, Wang, Huang, Wang, and Cao]{fang2023eva}
Fang, Y., Sun, Q., Wang, X., Huang, T., Wang, X., and Cao, Y.
\newblock Eva-02: A visual representation for neon genesis.
\newblock \emph{arXiv preprint arXiv:2303.11331}, 2023.

\bibitem[Fayek \& Kumar(2020)Fayek and Kumar]{fayek2020large}
Fayek, H.~M. and Kumar, A.
\newblock Large scale audiovisual learning of sounds with weakly labeled data.
\newblock \emph{arXiv preprint arXiv:2006.01595}, 2020.

\bibitem[Gan et~al.(2020)Gan, Huang, Chen, Tenenbaum, and Torralba]{gan2020foley}
Gan, C., Huang, D., Chen, P., Tenenbaum, J.~B., and Torralba, A.
\newblock Foley music: Learning to generate music from videos.
\newblock In \emph{Computer Vision--ECCV 2020: 16th European Conference, Glasgow, UK, August 23--28, 2020, Proceedings, Part XI 16}, pp.\  758--775. Springer, 2020.

\bibitem[Garcia et~al.(2023)Garcia, Seetharaman, Kumar, and Pardo]{garcia2023vampnet}
Garcia, H.~F., Seetharaman, P., Kumar, R., and Pardo, B.
\newblock Vampnet: Music generation via masked acoustic token modeling.
\newblock \emph{arXiv preprint arXiv:2307.04686}, 2023.

\bibitem[Gemmeke et~al.(2017)Gemmeke, Ellis, Freedman, Jansen, Lawrence, Moore, Plakal, and Ritter]{gemmeke2017audio}
Gemmeke, J.~F., Ellis, D.~P., Freedman, D., Jansen, A., Lawrence, W., Moore, R.~C., Plakal, M., and Ritter, M.
\newblock Audio set: An ontology and human-labeled dataset for audio events.
\newblock In \emph{2017 IEEE international conference on acoustics, speech and signal processing (ICASSP)}, pp.\  776--780. IEEE, 2017.

\bibitem[Georgescu et~al.(2023)Georgescu, Fonseca, Ionescu, Lucic, Schmid, and Arnab]{georgescu2023audiovisual}
Georgescu, M.-I., Fonseca, E., Ionescu, R.~T., Lucic, M., Schmid, C., and Arnab, A.
\newblock Audiovisual masked autoencoders.
\newblock In \emph{Proceedings of the IEEE/CVF International Conference on Computer Vision}, pp.\  16144--16154, 2023.

\bibitem[Girdhar et~al.(2023)Girdhar, El-Nouby, Liu, Singh, Alwala, Joulin, and Misra]{girdhar2023imagebind}
Girdhar, R., El-Nouby, A., Liu, Z., Singh, M., Alwala, K.~V., Joulin, A., and Misra, I.
\newblock Imagebind: One embedding space to bind them all, 2023.

\bibitem[Gong et~al.(2021)Gong, Chung, and Glass]{gong2021ast}
Gong, Y., Chung, Y.-A., and Glass, J.
\newblock Ast: Audio spectrogram transformer.
\newblock \emph{arXiv preprint arXiv:2104.01778}, 2021.

\bibitem[Gong et~al.(2022{\natexlab{a}})Gong, Lai, Chung, and Glass]{gong2022ssast}
Gong, Y., Lai, C.-I., Chung, Y.-A., and Glass, J.
\newblock Ssast: Self-supervised audio spectrogram transformer.
\newblock In \emph{Proceedings of the AAAI Conference on Artificial Intelligence}, volume~36, pp.\  10699--10709, 2022{\natexlab{a}}.

\bibitem[Gong et~al.(2022{\natexlab{b}})Gong, Rouditchenko, Liu, Harwath, Karlinsky, Kuehne, and Glass]{gong2022contrastive}
Gong, Y., Rouditchenko, A., Liu, A.~H., Harwath, D., Karlinsky, L., Kuehne, H., and Glass, J.
\newblock Contrastive audio-visual masked autoencoder.
\newblock \emph{arXiv preprint arXiv:2210.07839}, 2022{\natexlab{b}}.

\bibitem[He et~al.(2022)He, Chen, Xie, Li, Doll{\'a}r, and Girshick]{he2022masked}
He, K., Chen, X., Xie, S., Li, Y., Doll{\'a}r, P., and Girshick, R.
\newblock Masked autoencoders are scalable vision learners.
\newblock In \emph{Proceedings of the IEEE/CVF conference on computer vision and pattern recognition}, pp.\  16000--16009, 2022.

\bibitem[Hershey et~al.(2017)Hershey, Chaudhuri, Ellis, Gemmeke, Jansen, Moore, Plakal, Platt, Saurous, Seybold, et~al.]{hershey2017cnn}
Hershey, S., Chaudhuri, S., Ellis, D.~P., Gemmeke, J.~F., Jansen, A., Moore, R.~C., Plakal, M., Platt, D., Saurous, R.~A., Seybold, B., et~al.
\newblock Cnn architectures for large-scale audio classification.
\newblock In \emph{2017 ieee international conference on acoustics, speech and signal processing (icassp)}, pp.\  131--135. IEEE, 2017.

\bibitem[Ho \& Salimans(2022)Ho and Salimans]{ho2022classifier}
Ho, J. and Salimans, T.
\newblock Classifier-free diffusion guidance.
\newblock \emph{arXiv preprint arXiv:2207.12598}, 2022.

\bibitem[Huang et~al.(2022{\natexlab{a}})Huang, Sharma, Xu, Ryali, Fan, Li, Li, Ghosh, Malik, and Feichtenhofer]{huang2022mavil}
Huang, P.-Y., Sharma, V., Xu, H., Ryali, C., Fan, H., Li, Y., Li, S.-W., Ghosh, G., Malik, J., and Feichtenhofer, C.
\newblock Mavil: Masked audio-video learners.
\newblock \emph{arXiv preprint arXiv:2212.08071}, 2022{\natexlab{a}}.

\bibitem[Huang et~al.(2022{\natexlab{b}})Huang, Xu, Li, Baevski, Auli, Galuba, Metze, and Feichtenhofer]{huang2022masked}
Huang, P.-Y., Xu, H., Li, J., Baevski, A., Auli, M., Galuba, W., Metze, F., and Feichtenhofer, C.
\newblock Masked autoencoders that listen.
\newblock \emph{Advances in Neural Information Processing Systems}, 35:\penalty0 28708--28720, 2022{\natexlab{b}}.

\bibitem[Huang et~al.(2023)Huang, Xu, Li, Baevski, Auli, Galuba, Metze, and Feichtenhofer]{huang2023masked}
Huang, P.-Y., Xu, H., Li, J., Baevski, A., Auli, M., Galuba, W., Metze, F., and Feichtenhofer, C.
\newblock Masked autoencoders that listen, 2023.

\bibitem[Iashin \& Rahtu(2021)Iashin and Rahtu]{iashin2021taming}
Iashin, V. and Rahtu, E.
\newblock Taming visually guided sound generation, 2021.

\bibitem[Jaegle et~al.(2021)Jaegle, Gimeno, Brock, Vinyals, Zisserman, and Carreira]{jaegle2021perceiver}
Jaegle, A., Gimeno, F., Brock, A., Vinyals, O., Zisserman, A., and Carreira, J.
\newblock Perceiver: General perception with iterative attention.
\newblock In \emph{International conference on machine learning}, pp.\  4651--4664. PMLR, 2021.

\bibitem[Kilgour et~al.(2018)Kilgour, Zuluaga, Roblek, and Sharifi]{kilgour2018fr}
Kilgour, K., Zuluaga, M., Roblek, D., and Sharifi, M.
\newblock Fr$\backslash$'echet audio distance: A metric for evaluating music enhancement algorithms.
\newblock \emph{arXiv preprint arXiv:1812.08466}, 2018.

\bibitem[Kim et~al.(2013)Kim, Lee, and Provost]{kim2013deep}
Kim, Y., Lee, H., and Provost, E.~M.
\newblock Deep learning for robust feature generation in audiovisual emotion recognition.
\newblock In \emph{2013 IEEE international conference on acoustics, speech and signal processing}, pp.\  3687--3691. IEEE, 2013.

\bibitem[Koutini et~al.(2021)Koutini, Schl{\"u}ter, Eghbal-Zadeh, and Widmer]{koutini2021efficient}
Koutini, K., Schl{\"u}ter, J., Eghbal-Zadeh, H., and Widmer, G.
\newblock Efficient training of audio transformers with patchout.
\newblock \emph{arXiv preprint arXiv:2110.05069}, 2021.

\bibitem[Kreuk et~al.(2022)Kreuk, Synnaeve, Polyak, Singer, D{\'e}fossez, Copet, Parikh, Taigman, and Adi]{kreuk2022audiogen}
Kreuk, F., Synnaeve, G., Polyak, A., Singer, U., D{\'e}fossez, A., Copet, J., Parikh, D., Taigman, Y., and Adi, Y.
\newblock Audiogen: Textually guided audio generation.
\newblock \emph{arXiv preprint arXiv:2209.15352}, 2022.

\bibitem[Kumar et~al.(2023)Kumar, Seetharaman, Luebs, Kumar, and Kumar]{kumar2023high}
Kumar, R., Seetharaman, P., Luebs, A., Kumar, I., and Kumar, K.
\newblock High-fidelity audio compression with improved rvqgan.
\newblock \emph{arXiv preprint arXiv:2306.06546}, 2023.

\bibitem[Li et~al.(2023{\natexlab{a}})Li, Li, Savarese, and Hoi]{li2023blip}
Li, J., Li, D., Savarese, S., and Hoi, S.
\newblock Blip-2: Bootstrapping language-image pre-training with frozen image encoders and large language models.
\newblock \emph{arXiv preprint arXiv:2301.12597}, 2023{\natexlab{a}}.

\bibitem[Li et~al.(2023{\natexlab{b}})Li, Chang, Mishra, Zhang, Katabi, and Krishnan]{li2023mage}
Li, T., Chang, H., Mishra, S., Zhang, H., Katabi, D., and Krishnan, D.
\newblock Mage: Masked generative encoder to unify representation learning and image synthesis.
\newblock In \emph{Proceedings of the IEEE/CVF Conference on Computer Vision and Pattern Recognition}, pp.\  2142--2152, 2023{\natexlab{b}}.

\bibitem[Liu et~al.(2024{\natexlab{a}})Liu, Dong, and Zhang]{liu2024tackling}
Liu, X., Dong, Z., and Zhang, P.
\newblock Tackling data bias in music-avqa: Crafting a balanced dataset for unbiased question-answering.
\newblock In \emph{Proceedings of the IEEE/CVF Winter Conference on Applications of Computer Vision (WACV)}, pp.\  4478--4487, January 2024{\natexlab{a}}.

\bibitem[Liu et~al.(2024{\natexlab{b}})Liu, Su, and Shlizerman]{Liu_2024_WACV}
Liu, X., Su, K., and Shlizerman, E.
\newblock Let the beat follow you - creating interactive drum sounds from body rhythm.
\newblock In \emph{Proceedings of the IEEE/CVF Winter Conference on Applications of Computer Vision (WACV)}, pp.\  7187--7197, January 2024{\natexlab{b}}.

\bibitem[Luo et~al.(2023)Luo, Yan, Hu, and Zhao]{luo2023diff}
Luo, S., Yan, C., Hu, C., and Zhao, H.
\newblock Diff-foley: Synchronized video-to-audio synthesis with latent diffusion models.
\newblock \emph{arXiv preprint arXiv:2306.17203}, 2023.

\bibitem[Ma et~al.(2020)Ma, Zeng, McDuff, and Song]{ma2020active}
Ma, S., Zeng, Z., McDuff, D., and Song, Y.
\newblock Active contrastive learning of audio-visual video representations.
\newblock \emph{arXiv preprint arXiv:2009.09805}, 2020.

\bibitem[Mehri et~al.(2016)Mehri, Kumar, Gulrajani, Kumar, Jain, Sotelo, Courville, and Bengio]{mehri2016samplernn}
Mehri, S., Kumar, K., Gulrajani, I., Kumar, R., Jain, S., Sotelo, J., Courville, A., and Bengio, Y.
\newblock Samplernn: An unconditional end-to-end neural audio generation model.
\newblock \emph{arXiv preprint arXiv:1612.07837}, 2016.

\bibitem[Mei et~al.(2023)Mei, Nagaraja, Lan, Ni, Chang, Shi, and Chandra]{mei2023foleygen}
Mei, X., Nagaraja, V., Lan, G.~L., Ni, Z., Chang, E., Shi, Y., and Chandra, V.
\newblock Foleygen: Visually-guided audio generation.
\newblock \emph{arXiv preprint arXiv:2309.10537}, 2023.

\bibitem[Morgado et~al.(2021)Morgado, Vasconcelos, and Misra]{morgado2021audio}
Morgado, P., Vasconcelos, N., and Misra, I.
\newblock Audio-visual instance discrimination with cross-modal agreement.
\newblock In \emph{Proceedings of the IEEE/CVF Conference on Computer Vision and Pattern Recognition}, pp.\  12475--12486, 2021.

\bibitem[Nagrani et~al.(2021)Nagrani, Yang, Arnab, Jansen, Schmid, and Sun]{nagrani2021attention}
Nagrani, A., Yang, S., Arnab, A., Jansen, A., Schmid, C., and Sun, C.
\newblock Attention bottlenecks for multimodal fusion.
\newblock \emph{Advances in Neural Information Processing Systems}, 34:\penalty0 14200--14213, 2021.

\bibitem[Ngiam et~al.(2011)Ngiam, Khosla, Kim, Nam, Lee, and Ng]{ngiam2011multimodal}
Ngiam, J., Khosla, A., Kim, M., Nam, J., Lee, H., and Ng, A.~Y.
\newblock Multimodal deep learning.
\newblock In \emph{Proceedings of the 28th international conference on machine learning (ICML-11)}, pp.\  689--696, 2011.

\bibitem[Owens et~al.(2016)Owens, Isola, McDermott, Torralba, Adelson, and Freeman]{owens2016visually}
Owens, A., Isola, P., McDermott, J., Torralba, A., Adelson, E.~H., and Freeman, W.~T.
\newblock Visually indicated sounds.
\newblock In \emph{Proceedings of the IEEE conference on computer vision and pattern recognition}, pp.\  2405--2413, 2016.

\bibitem[Piczak(2015)]{piczak2015esc}
Piczak, K.~J.
\newblock Esc: Dataset for environmental sound classification.
\newblock In \emph{Proceedings of the 23rd ACM international conference on Multimedia}, pp.\  1015--1018, 2015.

\bibitem[Press et~al.(2021)Press, Smith, and Lewis]{press2021train}
Press, O., Smith, N.~A., and Lewis, M.
\newblock Train short, test long: Attention with linear biases enables input length extrapolation.
\newblock \emph{arXiv preprint arXiv:2108.12409}, 2021.

\bibitem[Radford et~al.(2021)Radford, Kim, Hallacy, Ramesh, Goh, Agarwal, Sastry, Askell, Mishkin, Clark, et~al.]{radford2021learning}
Radford, A., Kim, J.~W., Hallacy, C., Ramesh, A., Goh, G., Agarwal, S., Sastry, G., Askell, A., Mishkin, P., Clark, J., et~al.
\newblock Learning transferable visual models from natural language supervision.
\newblock In \emph{International conference on machine learning}, pp.\  8748--8763. PMLR, 2021.

\bibitem[Sheffer \& Adi(2023)Sheffer and Adi]{sheffer2023hear}
Sheffer, R. and Adi, Y.
\newblock I hear your true colors: Image guided audio generation.
\newblock In \emph{ICASSP 2023-2023 IEEE International Conference on Acoustics, Speech and Signal Processing (ICASSP)}, pp.\  1--5. IEEE, 2023.

\bibitem[Su et~al.(2020{\natexlab{a}})Su, Liu, and Shlizerman]{Su_2020_CVPR}
Su, K., Liu, X., and Shlizerman, E.
\newblock Predict \& cluster: Unsupervised skeleton based action recognition.
\newblock In \emph{Proceedings of the IEEE/CVF Conference on Computer Vision and Pattern Recognition (CVPR)}, June 2020{\natexlab{a}}.

\bibitem[Su et~al.(2020{\natexlab{b}})Su, Liu, and Shlizerman]{su2020audeo}
Su, K., Liu, X., and Shlizerman, E.
\newblock Audeo: Audio generation for a silent performance video.
\newblock \emph{Advances in Neural Information Processing Systems}, 33:\penalty0 3325--3337, 2020{\natexlab{b}}.

\bibitem[Su et~al.(2020{\natexlab{c}})Su, Liu, and Shlizerman]{su2020multi}
Su, K., Liu, X., and Shlizerman, E.
\newblock Multi-instrumentalist net: Unsupervised generation of music from body movements.
\newblock \emph{arXiv preprint arXiv:2012.03478}, 2020{\natexlab{c}}.

\bibitem[Su et~al.(2021)Su, Liu, and Shlizerman]{su2021does}
Su, K., Liu, X., and Shlizerman, E.
\newblock How does it sound?
\newblock \emph{Advances in Neural Information Processing Systems}, 34:\penalty0 29258--29273, 2021.

\bibitem[Su et~al.(2023{\natexlab{a}})Su, Li, Huang, Kuzmin, Lee, Donahue, Sha, Jansen, Wang, Verzetti, et~al.]{su2023v2meow}
Su, K., Li, J.~Y., Huang, Q., Kuzmin, D., Lee, J., Donahue, C., Sha, F., Jansen, A., Wang, Y., Verzetti, M., et~al.
\newblock V2meow: Meowing to the visual beat via music generation.
\newblock \emph{arXiv preprint arXiv:2305.06594}, 2023{\natexlab{a}}.

\bibitem[Su et~al.(2023{\natexlab{b}})Su, Qian, Shlizerman, Torralba, and Gan]{su2023physics}
Su, K., Qian, K., Shlizerman, E., Torralba, A., and Gan, C.
\newblock Physics-driven diffusion models for impact sound synthesis from videos.
\newblock In \emph{Proceedings of the IEEE/CVF Conference on Computer Vision and Pattern Recognition}, pp.\  9749--9759, 2023{\natexlab{b}}.

\bibitem[Tong et~al.(2022)Tong, Song, Wang, and Wang]{tong2022videomae}
Tong, Z., Song, Y., Wang, J., and Wang, L.
\newblock Videomae: Masked autoencoders are data-efficient learners for self-supervised video pre-training, 2022.

\bibitem[Wang et~al.(2020)Wang, Tran, and Feiszli]{wang2020makes}
Wang, W., Tran, D., and Feiszli, M.
\newblock What makes training multi-modal classification networks hard?
\newblock In \emph{Proceedings of the IEEE/CVF conference on computer vision and pattern recognition}, pp.\  12695--12705, 2020.

\bibitem[Wang et~al.(2022)Wang, Bao, Dong, Bjorck, Peng, Liu, Aggarwal, Mohammed, Singhal, Som, et~al.]{wang2022image}
Wang, W., Bao, H., Dong, L., Bjorck, J., Peng, Z., Liu, Q., Aggarwal, K., Mohammed, O.~K., Singhal, S., Som, S., et~al.
\newblock Image as a foreign language: Beit pretraining for all vision and vision-language tasks.
\newblock \emph{arXiv preprint arXiv:2208.10442}, 2022.

\bibitem[Warden(2018)]{warden2018speech}
Warden, P.
\newblock Speech commands: A dataset for limited-vocabulary speech recognition.
\newblock \emph{arXiv preprint arXiv:1804.03209}, 2018.

\bibitem[Xu et~al.(2016)Xu, Mei, Yao, and Rui]{xu2016msr}
Xu, J., Mei, T., Yao, T., and Rui, Y.
\newblock Msr-vtt: A large video description dataset for bridging video and language.
\newblock In \emph{Proceedings of the IEEE conference on computer vision and pattern recognition}, pp.\  5288--5296, 2016.

\bibitem[Yu et~al.(2022)Yu, Wang, Vasudevan, Yeung, Seyedhosseini, and Wu]{yu2022coca}
Yu, J., Wang, Z., Vasudevan, V., Yeung, L., Seyedhosseini, M., and Wu, Y.
\newblock Coca: Contrastive captioners are image-text foundation models.
\newblock \emph{arXiv preprint arXiv:2205.01917}, 2022.

\bibitem[Yu et~al.(2023)Yu, Cheng, Sohn, Lezama, Zhang, Chang, Hauptmann, Yang, Hao, Essa, et~al.]{yu2023magvit}
Yu, L., Cheng, Y., Sohn, K., Lezama, J., Zhang, H., Chang, H., Hauptmann, A.~G., Yang, M.-H., Hao, Y., Essa, I., et~al.
\newblock Magvit: Masked generative video transformer.
\newblock In \emph{Proceedings of the IEEE/CVF Conference on Computer Vision and Pattern Recognition}, pp.\  10459--10469, 2023.

\bibitem[Zhou et~al.(2018)Zhou, Wang, Fang, Bui, and Berg]{zhou2018visual}
Zhou, Y., Wang, Z., Fang, C., Bui, T., and Berg, T.~L.
\newblock Visual to sound: Generating natural sound for videos in the wild.
\newblock In \emph{Proceedings of the IEEE conference on computer vision and pattern recognition}, pp.\  3550--3558, 2018.

\bibitem[Zhu et~al.(2023)Zhu, Lin, Ning, Yan, Cui, Wang, Pang, Jiang, Zhang, Li, et~al.]{zhu2023languagebind}
Zhu, B., Lin, B., Ning, M., Yan, Y., Cui, J., Wang, H., Pang, Y., Jiang, W., Zhang, J., Li, Z., et~al.
\newblock Languagebind: Extending video-language pretraining to n-modality by language-based semantic alignment.
\newblock \emph{arXiv preprint arXiv:2310.01852}, 2023.

\end{thebibliography}
\bibliographystyle{icml2024}

%%%%%%%%%%%%%%%%%%%%%%%%%%%%%%%%%%%%%%%%%%%%%%%%%%%%%%%%%%%%%%%%%%%%%%%%%%%%%%%
%%%%%%%%%%%%%%%%%%%%%%%%%%%%%%%%%%%%%%%%%%%%%%%%%%%%%%%%%%%%%%%%%%%%%%%%%%%%%%%
% APPENDIX
%%%%%%%%%%%%%%%%%%%%%%%%%%%%%%%%%%%%%%%%%%%%%%%%%%%%%%%%%%%%%%%%%%%%%%%%%%%%%%%
%%%%%%%%%%%%%%%%%%%%%%%%%%%%%%%%%%%%%%%%%%%%%%%%%%%%%%%%%%%%%%%%%%%%%%%%%%%%%%%
\newpage
\appendix
\onecolumn
\section{Human Evaluation Details}
\label{human_eval}
For the human study, we conducted random sampling of 150 distinct video examples from the class-balanced VGGSound test subset, which was used in the retrieval task. This subset consists of 5 videos per class, totaling 1545 videos. In the rating process, we presented each pair of videos to human raters and requested them to perform a side-by-side comparison of the audio generated by the baseline models and the VAB model. Raters were asked to assess the two videos in terms of two criteria: 1) overall quality and 2) visual relevance. Ratings were provided on a 5-point Likert scale (Strong Preference Video 1, Weak Preference Video 1, No Preference, Weak Preference Video 2, Strong Preference Video 2). To facilitate the surveys, we leveraged the \href{https://toloka.ai/}{Toloka platform}, with each video pair being assessed by three individuals. The compensation for each assignment is set at $\$$0.05. In an effort to prevent potential perceptual biases, no background information regarding the survey or our approach was disclosed to the participants. In total, we collected a set of 450 ratings for each comparison. The detailed results are shown in the Figure~\ref{fig:human_eval}.

\section{Experimental Details}
In this section, we offer details of our experimental settings encompassing data pre-processing, model architecture details, data augmentation, hyper-parameters configurations of pre-training and fine-tuning. A comprehensive summary of the training configuration can be found in accompanying Table~\ref{tab:train_config}. Additionally, our code and the pre-trained models will be made accessible for reference.

\subsection{Data Pre-processing}
In this work, we were able to collect a total of 1573485 videos for AS-2M training dataset. It's worth noting that, due to data availability constraints, our dataset size is smaller than those employed in related works such as MaViL~\cite{huang2022mavil}(2.01 million) and CAV-MAE~\cite{gong2022contrastive}(1.79 million). For VGGSound dataset, we obtained 177K training videos. By combining data from AudioSet and VGGSound, our pre-training dataset encompasses approximately 1.74 million training instances. We extracted audio tracks from the videos and resampled them to 16Khz. For AudioSet videos, we extracted video frames at a rate of 1 fps. For VGGSound videos, we initially extracted video frames at 5 fps so that we can augment the data by randomly selecting 1 frame for each second during training.

To extract audio tokens, we explored the utilization of open-source pre-trained \href{https://github.com/descriptinc/descript-audio-codec/tree/main}{DAC} and \href{https://github.com/facebookresearch/audiocraft/tree/main}{Encodec} tokens. The DAC contains 12 codebooks, each containing 1024 tokens, while Encodec has 4 codebooks, each comprising 2048 tokens. For each audio data sample, we adjusted the length to 10 seconds through cutting or padding and subsequently inferred the corresponding audio tokens. It's worth noting that both DAC and Encodec tokens operate at a rate of 50Hz. Regarding visual features, we harnessed the image encoder from eva02-CLIP-L~\cite{fang2023eva} to extract CLIP embeddings. These embeddings possess a feature dimension $768$ and are inferred for each frame. In cases where a video was shorter than 10 seconds, we padded the embedding of the last available frame. Both the extraction of audio tokens and CLIP embedding were performed prior to the VAB model training. During the VAB pre-training phase, the model received inputs consisting solely of audio tokens and CLIP embeddings.

\subsection{Model Architecture}
\label{model_arch}
We conducted experiments with VAB using both Encodec and DAC tokens. We have presented and reported the results of VAB-Encodec and VAB-DAC in our main paper. Additionally, we introduced a smaller model variant known as VAB-DAC-Test, designed to facilitate more efficient ablation studies and analyses, considering our limited computational resources. For DAC tokens, we also trained an additional Coarse2Fine module tailored for audio generation purposes. The architecture and hyperparameters of VAB-DAC-Coarse2Fine closely align with the one utilized in VampNet~\cite{garcia2023vampnet}. Detailed information about the model architectures can be found in the accompanying table~\ref{tab:model_arch}. 
\begin{table*}[th]
\footnotesize
\centering
\begin{tabular}{lcccc}
& VAB-Encodec & VAB-DAC & VAB-DAC-Test & VAB-DAC-Corase2Fine\\
\toprule
Visual Features Length & 10 & 10 & 10 & 10\\
Audio Tokens Length & 500 & 500 & 500 & 500\\
Number of Codebooks & 4 & 4 & 4 & 12\\
Codebook Dimension & 128 & 8 & 8 & 8\\
Number of Tokens & 2048 & 1024 & 1024 & 1024\\
Hidden Dimension & 1024 & 1024 & 768 & 1280\\
Total Layers & 24 & 24 & 20 & 16\\
Expert Layers & 12 & 12 & 12 & N/A\\
Attention Heads & 16 & 16 & 8 & 20\\
Positional Encoding & \multicolumn{4}{c}{Alibi Positional Embedding~\cite{press2021train}}\\
Number of Prediction Heads & 4 & 4 & 4 & 8\\
\bottomrule
\end{tabular}
\caption{Model Architecture}
\label{tab:model_arch}
\end{table*}

\subsection{Data Augmentation}
Since we pre-processed all audio tokens and visual features, we did not perform data augmentation on raw images and audio spectrogram levels during training. In fact, we utilize three types of augmentation during the pre-training phase. The primary augmentation technique is known as \textbf{temporal mixup}. Specifically, we begin by randomly selecting two pairs of data from the training set. Subsequently, we randomly choose a breaking point within the 0 to 10-second range. The first segment is then used for the first pair, while the second segment is assigned to the second pair. This augmentation technique, introduced in Diff-Foley~\cite{luo2023diff}, has proven effective in enhancing both generation capabilities and learning cross-modal relationships. Additionally, we apply \textbf{temporal rolling} to both audio tokens and visual features, akin to the rolling of a spectrogram in the temporal axis. Finally, to facilitate classifier-free guidance during inference, we drop out visual features with a probability of 0.1 during the pre-training phase. During fine-tuning for all tasks, we only employ the first two augmentations.

\subsection{Training Configuration and Hyper-parameters}
\label{train_config}
In this section, we provide an overview of the VAB pre-training and fine-tuning configurations applied to all datasets. Due to the computational resource constraints, we conducted all our training experiments using only 1 A100 GPU (80GB) and 1 A100 GPU (40GB). While these constraints limit the batch sizes, potentially impacting performance, it's worth noting that our training requirements are considerably more cost-effective compared to previous works. A comprehensive breakdown of these configurations is presented in the Table~\ref{tab:train_config}.
\begin{table*}[th]
\footnotesize
\centering
\begin{tabular}{lccccccc}
& Pre-training & Fine-tuning (Contrastive) & \multicolumn{5}{c}{Fine-tuning (Classification)}\\
\toprule
Model & \multicolumn{7}{c}{VAB-Encodec/VAB-DAC}\\
Configuration & AS-2M + VGGSound & AS-2M + VGGSound & AS-2M & AS-20K & VGGSound & ESC & SPC\\
Optimizer & \multicolumn{7}{c}{AdamW, $\beta_1=0.9, \beta_2=0.95$, Weight decay 1e-5}\\
Base learning rate & 2e-4 & 2e-4 & 1e-4 & 1e-4 & 1e-4 & 1e-4 & 1e-4\\
Learning rate schedule & \multicolumn{7}{c}{Half-cycle cosine decay}\\
Warm-up steps & 100K & 100K & 0 & 0 & 0 & 0 & 0 \\
Epochs & 17/9 & 14/15 & 2 & 10 & 5 & 50 & 10/5\\
Batch size & 60 & 108 & 60 & 60 & 48 & 60 & 60\\
GPU & 1 A100 (80G) & 1 A100 (80G) & \multicolumn{5}{c}{1A100 (40G)}\\
Temporal Mixup Prob. & \multicolumn{7}{c}{0.5}\\
Temporal Rolling Prob. & 0.1 & 0.1 & 0.1 & 0.1 & 0.1 & 1.0 & 1.0\\
Class Balanced Sampling & N/A & N/A & True & False & True & False & False\\
Label Smooth & \multicolumn{7}{c}{0.1}\\
Multilabel & N/A & N/A & True & True & False & False & False\\
Loss Function & CE & CE & BCE & BCE & BCE & BCE & BCE \\

\bottomrule
\end{tabular}
\caption{Training Configuration}
\label{tab:train_config}
\end{table*}

\section{Ablation Studies}
\subsection{Pre-training v.s. From Scratch}
\label{from_scratch}
We use the VAB-DAC-Test to juxtapose the outcomes of training VGGSound classification from scratch against our fine-tuning results. The results are shown in the first line of Table~\ref{tab:ablation_scratch_expert}. It becomes evident that a notable disparity exists in both single and multi-modal classification, underscoring the effectiveness of our pre-training approach.
\begin{table*}[th]
\centering
\begin{tabular}{lccc}
\toprule
& \multicolumn{3}{c}{\underline{VGGSound (Acc.)}$\uparrow$} \\
Method & V+A & V & A \\
\midrule
VAB-DAC-Test From Scratch & 54.7 & 52.8 & 38.6\\
VAB-DAC-Test w.o. experts & 42.5 & 48.9 & 41.0\\
VAB-DAC-Test & 62.4 & 53.2 & 44.5\\
\bottomrule
\end{tabular}
\caption{VGGSound classification comparisons for VAB-DAC-Test training from scratch and without modal specific experts}
\label{tab:ablation_scratch_expert}
\end{table*}

\subsection{Modal-specific Experts}
\label{expert}
We further delve into the influence of incorporating the experts module in the first $N_1=12$ layers. We conducted both masked audio tokens prediction pre-training and fine-tuning steps using the same methodology employed in the original VAB model. The results are presented in the second line of Table~\ref{tab:ablation_scratch_expert}. We found that in the absence of the experts module, the results reveal a deterioration in audio-visual joint classification and the performance is even worse than the visual-only outcomes. This phenomenon can likely be attributed to the longer audio tokens (500) than visual features (10) dominating the model's features, hindering the effective fusion of inputs from both modalities.

\subsection{Choice of Pre-training task: Masked Token Prediction v.s. Contrastive}
\label{choice}
Since both masked token prediction and contrastive learning are self-supervised tasks, understanding the optimal training order of applying these tasks becomes an intriguing endeavor. Prior research, such as CAV-MAE~\cite{gong2022contrastive} and MaViL~\cite{huang2022mavil}, effectively utilized both MAE loss and Contrastive loss during training. However, our experimentation revealed that simultaneous application of both losses during the pre-training phase led to failure of convergence. This challenge may stem from the multiple forward passes to the same weights for various targets, resulting in conflicting gradient directions. Therefore, we embark on an exploration to discern which order could prove more advantageous. To achieve this, we trained the same VAB-DAC-Test model in two directions. In the first case, we initiated pre-training of the VAB-DAC-Test with the masked audio token prediction task and then fine-tuned it with contrastive loss. In the second case, we start the pre-training of the VAB-DAC-Test with contrastive loss and followed it with fine-tuning for masked audio token prediction. Both cases employed identical hyperparameters and data. The comparison of retrieval and generation performances are presented in Table~\ref{tab:ablation_order}. We discovered that both orders actually yielded similar losses upon convergence and exhibited comparable performance on both tasks. Notably, the Masked Prediction $\rightarrow$ Contrastive direction required fewer total training epochs to converge than the Contrastive $\rightarrow$ Masked Prediction direction. This observation can be attributed to the inherent complexity of the masked audio token prediction task, which proves more challenging than identifying audio-visual pairs, especially considering our limited batch size of 108 for contrastive learning. Therefore, we ultimately opted for the masked token prediction task as the final pre-training task.
\begin{table*}[th]
\footnotesize
\centering
\begin{tabular}{lcccccccc}
\toprule
Training Order (training epochs) & \multicolumn{1}{c}{Generation} & \multicolumn{3}{c}{VGGSound V$\rightarrow$A Retrieval} & \multicolumn{3}{c}{VGGSound A$\rightarrow$V Retrieval}\\
& FAD $\downarrow$ & R@1 &  R@5 & R@10 & R@1 &  R@5 & R@10\\
\midrule
Contrastive (29) $\rightarrow$ Masked Prediction (9) & 4.14 & 27.7 & 56.4 & 67.5 & 32.5 & 60.8 & 72.3\\
Masked Prediction (9) $\rightarrow$ Contrastive (15) & 4.18 & 28.2 & 56.7 & 68.5 & 30.5 & 57.6 & 71.2\\
\bottomrule
\end{tabular}
\caption{Retrieval and generation performances under different training orders.}
\label{tab:ablation_order}
\end{table*}
\subsection{Effects of Contrastive Fine-tuning on Classification Tasks}
\label{pretrain_effect}
In this section, we delve into the question of whether contrastive fine-tuning can further enhance the classification tasks. The results presented in Table~\ref{tab:ablation_pretrain} clearly demonstrate that models subjected to contrastive fine-tuning consistently outperform their counterparts that underwent pre-training based solely on masked audio token prediction.
\begin{table*}[th]
\centering
\begin{tabular}{lccccccccccc}
\toprule
& \multicolumn{3}{c}{\underline{VGGSound (Acc.)}$\uparrow$} & \multicolumn{3}{c}{\underline{AS-2M (mAP)}$\uparrow$} &\multicolumn{3}{c}{\underline{AS-20K (mAP)}$\uparrow$}  \\
Method & V+A & V & A & V+A & V & A & V+A & V & A \\
\midrule
VAB-DAC-Test w.o. contrastive ft & 56.2 & 53.5 & 41.0 & 41.6 & 30.4 & 30.1 & 24.4 & 25.9 & 15.9\\
VAB-DAC-Test w. contrastive ft & 62.4 & 53.2 & 44.5 & 42.6 & 30.2& 31.1 & 37.7 & 27.7 & 22.2\\
VAB-DAC w.o. contrastive ft & 60.4 & 55.2 & 42.1 & 42.4 & 31.3 & 31.6 & 34.6 & 28.2 & 16.2\\
VAB-DAC w. contrastive ft & 63.9 & 55.4 & 48.2 & 47.0 & 33.3 & 36.2 & 38.9 & 28.3 & 28.8\\
VAB-Encodec w.o. contrastive ft & 62.9 & 54.6 & 47.4 & 46.7 & 33.0 & 35.7 & 36.0 & 27.5 & 24.0\\
VAB-Encodec w. contrastive ft & 65.2 & 55.1 & 51.3 & 47.7 & 33.5 & 38.6 & 38.7 & 29.0 & 29.0\\
\bottomrule
\end{tabular}
\caption{Impact of contrastive fine-tuning on classification tasks.}
\label{tab:ablation_pretrain}
\end{table*}

\subsection{Masking Strategy}
\label{mask_strategy}
In this section, we explore how the masking ratio employed during pre-training influences the classification and generation performances. We conducted training with VAB-DAC-Test using three different masking ratios. Our findings revealed that employing a larger portion of masks ($75\%$) during pre-training led to an improvement in VGGSound classification performance, a trend consistent with observations made in prior MAE-based approaches. However, it's essential to note that this approach resulted in suboptimal audio quality for video-to-audio generation. Conversely, opting for a smaller masking proportion ($35\%$) failed to effectively learn both representation and generation. The detailed results are presented in Table~\ref{tab:masked_strategy}.
\begin{table*}[h]
\centering
\begin{tabular}{lcc}
\toprule
Masking Ratio & \underline{V + A Classification (Acc.)}$\uparrow$ & \underline{Generation FAD}$\downarrow$\\
mean = 0.35, std = 0.25 & 55.2 & 6.04\\
mean = 0.55, std = 0.25 & 56.2 & 5.30\\
mean = 0.75, std = 0.25 & 57.2 & 6.55\\
\bottomrule
\end{tabular}
\caption{VAB-DAC-Test performances on VGGSound classification and generation using different masking ratio}
\label{tab:masked_strategy}
\end{table*}

\subsection{Effect of Visual Encoders}
\label{visual_encoders}
Furthermore, we discuss the effect of using different visual encoders to extract features from video on the VAB model performance. We conducted visual encoder ablation study by replacing our visual encoder with MAE, same as CAV-MAE. We extract the frame-level features of all the video data (with same fps) by extracting the “[CLS]” token from MAE’s pretrained ViT-Large (1024 dims) encoder, and use them in replacement of eva-CLIP features to retrain our model. To ablate, we adopt the VAB-DAC-Test model configuration as described in Table ~\ref{tab:model_arch} in Appendix \ref{model_arch}, and name it VAB-DAC-Test (MAE). Following the same pretraining and contrastive fine-tuning procedure, we compare VAB-DAC-Test (MAE) with our VAB-DAC-Test w. contrastive ft (Table \ref{tab:ablation_scratch_expert}) on downstream tasks including classification, retrieval and audio generation tasks.

\textbf{Classification:} As shown in Table \ref{tab:visual_encoder_1}, VAB-DAC-Test with eva-CLIP as visual encoder consistently outperforms VAB-DAC-Test (MAE) across three datasets (VGGSound, AudioSet-20K and Audioset-2M) on Visual and Audio-Visual Classification tasks, while slightly falling behind in Audio Classification.

\textbf{Retrieval:} In Table \ref{tab:visual_encoder_2}, we show retrieval comparison results on the VGGSound dataset. As can be seen from the table, the performances of VAB-DAC-Test (MAE) in retrieval on both $V \rightarrow A$ and $A \rightarrow V$ fall behind ours.

\textbf{Audio Generation:} When it comes to audio generation, also as shown in the leftmost column of Table \ref{tab:visual_encoder_2}, our VAB-DAC-Test with eva-CLIP outperforms VAB-DAC-Test (MAE).

These results are within our expectation since MAE is a weaker visual encoder than eva-CLIP. Thereby, we justify our choice of using a strong visual encoder to enable better generalization on both generation and understanding tasks.

\begin{table}[h]
\centering
\begin{tabular}{lccccccccc}
\hline
 & \multicolumn{3}{c}{VGGSound (Acc)} & \multicolumn{3}{c}{AS-20K (mAP)} & \multicolumn{3}{c}{AS-2M (mAP)} \\ \cline{2-10} 
 & V+A & V & A & V+A & V & A & V+A & V & A \\ \hline
VAB-DAC-Test (MAE) & 55.9 & 39.1 & \textbf{46.2} & 31.2 & 15.9 & \textbf{22.4} & 39.0 & 21.4 & \textbf{31.2} \\
VAB-DAC-Test (Ours) & \textbf{62.4} & \textbf{53.2} & 44.5 & \textbf{37.7} & \textbf{27.7} & 22.2 & \textbf{42.6} & \textbf{30.2} & 31.1 \\ \hline
\end{tabular}
\caption{Visual encoder ablation study (MAE vs. eva-CLIP) for classification task}
\label{tab:visual_encoder_1}
\end{table}

\begin{table}[h]
\centering
\begin{tabular}{lccccccc}
\hline
 & \multicolumn{1}{c}{Generation (FAD)} & \multicolumn{3}{c}{VGGSound V$\rightarrow$A Retrieval} & \multicolumn{3}{c}{VGGSound A$\rightarrow$V Retrieval} \\ \cline{2-8} 
 &  & R@1 & R@5 & R@10 & R@1 & R@5 & R@10 \\ \hline
VAB-DAC-Test (MAE) & 4.93 & 18.4 & 40.1 & 52.3 & 20.6 & 46.6 & 57.6 \\
VAB-DAC-Test (Ours) & \textbf{4.18} & \textbf{28.2} & \textbf{56.7} & \textbf{68.5} & \textbf{30.5} & \textbf{57.6} & \textbf{71.2} \\ \hline
\end{tabular}
\caption{Visual encoder ablation study (MAE v.s. eva-CLIP) for generation and retrieval tasks}
\label{tab:visual_encoder_2}
\end{table}

\section{Video-to-Audio Generation Analysis}
\label{gen_analysis}
\subsection{Upperbound of Encodec and DAC tokens}
\label{upperbound}
As Encodec and DAC tokens involve lossy compression, it is imperative to ascertain the upper bound of the audio generation quality. to achieve this, we initiated an evaluation of the FAD scores for the reconstructed audio obtained from the ground truth Encodec and DAC tokens of the VGGSound test set. In Table~\ref{tab:encodec_dac}, our findings revealed that the DAC tokens exhibited a better FAD score when compared to Encodec tokens. However, it's important to note that DAC, employing a larger number of codebooks (12 levels), necessitates token modeling to be trained in two stages. Consequently, we trained the DAC-Coarse2Fine model, following the Vampnet~\cite{garcia2023vampnet}, and assessed the FAD score while conditioning it on the first four levels of ground truth DAC tokens. The results elucidated that with the incorporation of the additional Coarse2Fine stage, the upper bound for DAC reached a FAD score of 1.32 with 24 decoding iterations and 1.28 with 36 decoding steps. While DAC exhibited a stronger upper bound compared to Encodec tokens, we observe that our VAB-Encodec model consistently outperformed VAB-DAC, as shown in Table~\ref{tab:gen}. We hypothesize that the reason behind this discrepancy may be attributed to the importance of the information lost in the coarse-level DAC tokens, which strongly influences the learning process on both representation and generation.
\begin{table*}[h]
\centering
\begin{tabular}{lcccc}
\toprule
& Encodec & DAC & DAC-Corase2Fine (iter 24) & DAC-Coarse2Fine (iter 36)\\
FAD Score & 1.86 & 0.89 & 1.32 & 1.28\\
\bottomrule
\end{tabular}
\caption{FAD score on the reconstruction of Encodec and DAC tokens for VGGSound test set.}
\label{tab:encodec_dac}
\end{table*}

\subsection{effects of Classifier-free Guidance Scale}
We conducted an exploration of different classifier-free guidance scales during inference. For this experiment, we used the VAB-DAC model, keeping the decoding steps fixed at $16$ and the masking temperature at $10.5$. The results are outlined in Table~\ref{tab:ablation_cfg}. We found that a classifier-free guidance value of $5$ yielded the best FAD scores in our case.
\begin{table*}[th]
\centering
\begin{tabular}{lccccc}
\toprule
Classifier-free Guidance Scale & None & 3 & 5 & 7 & 11\\
FAD & 5.93 & 3.50 & 3.24 & 3.39 & 4.18 \\
KLD & 3.45 & 2.90 & 2.84 & 2.83 & 2.82 \\
\bottomrule 
\end{tabular}
\caption{FAD and KLD score with various classifier-free guidance scale during inference}
\label{tab:ablation_cfg}
\end{table*}

\subsection{Effect of Masking Temperature}
\begin{table*}[h]
\centering
\begin{tabular}{lcccccc}
\toprule
Masking Temperature & 4.5 & 10.5 & 12.5 & 15.5 & 20.5 & 25.5\\
FAD & 5.75  & 3.1 & 2.98 & 2.83 & 2.74 & 2.69\\
KLD & 3.08 & 2.69 & 2.61 & 2.61 & 2.59 & 2.58\\
\bottomrule
\end{tabular}
\caption{FAD and KLD scores with various masking temperatures during inference}
\label{tab:ablation_temp}
\end{table*}
We also investigated how the temperature parameter used in the confidence score calculation affects the quality of the generated audio. In this study, we employed the VAB-Encodec model while keeping the classifier-free guidance scale fixed at $5$ and the decoding steps at $16$. Our findings indicated that higher temperature values resulted in the generation of higher-quality samples, as shown in Table~\ref{tab:ablation_temp}.

\subsection{Effect of Decoding Steps}
\begin{table*}[h]
\centering
\begin{tabular}{lccccc}
\toprule
Decoding Iterations & 8 & 16 & 36 & 48\\
FAD & 3.52 & 3.1 & 3.25 & 3.32\\
KLD & 2.67 & 2.65 & 2.69 & 2.71\\
\bottomrule
\end{tabular}
\caption{FAD and KLD scores with various decoding steps during inference}
\label{tab:ablation_iter}
\end{table*}
We examined how the number of decoding steps used in the generation affects the generated audio. We employed the VAB-Encodec model while keeping the classifier-free guidance scale fixed at $5$ and the masking temperature at $10.5$. Our findings indicated that employing about 16 iterations are sufficient to obtain reasonable audio quality while more and less might degrade the performances. The results are shown in Table~\ref{tab:ablation_iter}.

\section{Details of Generating Samples from Video-to-Audio Generation Baselines}
\label{baseline_gen}
In this section, we describe the process of generating samples from baseline video-to-audio generation baseline methods. For SpecVQGAN~\cite{iashin2021taming}, IM2WAV~\cite{sheffer2023hear}, and Diff-Foley~\cite{luo2023diff}, we utilized their readily available pre-trained models for conducting inference. However, it's worth noting that IM2WAV and Diff-Foley were originally designed for generating audio in shorter duration and not specifically for producing 10-second audio samples. To adapt these models for generating 10-second audio samples, we made the following adjustments. For IM2WAV, the original inference process generates 4-second audio. To obtain 10-second samples, we generated three non-overlapping 4-second audio segments and combined them to form the final 10-second sample. For Diff-Foley, it was originally designed to generate 8.2-second audio. To produce 10-second audio samples, we generated two segments: one from 0-8.2 seconds and another from 1.8-10 seconds within the same batch. Then, we padded the first sample with the last 1.8 seconds of the second sample to achieve the desired 10-second duration.

For FoleyGen~\cite{mei2023foleygen}, we replicate the model described in the paper, utilizing our extracted Encodec tokens. The model comprises a transformer decoder with 24 layers, each equipped with 16 heads, and featuring dimensions of 1024. We faithfully adhered to the same training strategies as outlined in the original paper and employed CLIP embeddings at a frame rate of 1 frame per second as conditional signals. Additionally, we maintained consistency with the original classifier-free guidance scale (cfg$=3.0$) and adopted a top-k sampling approach with $k=256$ for generating the final samples for comparison. It's worth noting that the FAD score reported in the original paper ($1.65$) is better than the upper bound achieved with our Encodec tokens ground truth ($1.86$). This suggests that the publicly available Encodec model may not be the optimal version. Even with the available Encodec tokens, our re-implemented FoleyGen model still surpasses all other methods in terms of FAD score, underscoring the robustness of this state-of-the-art autoregressive approach. However, FoleyGen, like other autoregressive methods, encounters challenges with lengthy generation times. In contrast, our VAB-Encodec significantly accelerates the generation process by 17 times while maintaining high audio quality.

\section{Limitations and Discussion}
While VAB successfully bridges the gap between audio-visual representation learning and video-to-audio generation, it is important to acknowledge its limitations. Firstly, due to storage constraints, we could only utilize visual features at a 1fps resolution across all our data, potentially resulting in the loss of crucial visual details for audio-visual events when compared to the 2fps resolution employed in MaViL. As a result, we do not anticipate generating perfectly synchronized audio for videos, though we believe it could be achievable with higher visual resolution. Additionally, our use of pre-processed frame-level CLIP embeddings has led to the omission of spatial information, which is vital for tasks such as audio-visual localization and audio-visual question answering. Nonetheless, we believe that with increased computational resources, it is possible to incorporate spatial features from CLIP image encoder into the model. Moreover, there still exists a notable performance gap in classification tasks and a more advanced approach needs to be considered to further enhance the representation learning with audio tokens. Lastly, VAB primarily focuses on video-to-audio generation. When considering audio-to-video or audio-to-image generation, two challenges emerged. First, computational resources become a limiting factor. Second, the image and video data in AudioSet and VGGSound contain significant noise and irrelevant information, making it challenging to learn meaningful image or video generation from audio cues.

% You can have as much text here as you want. The main body must be at most $8$ pages long.
% For the final version, one more page can be added.
% If you want, you can use an appendix like this one.  

% The $\mathtt{\backslash onecolumn}$ command above can be kept in place if you prefer a one-column appendix, or can be removed if you prefer a two-column appendix.  Apart from this possible change, the style (font size, spacing, margins, page numbering, etc.) should be kept the same as the main body.
%%%%%%%%%%%%%%%%%%%%%%%%%%%%%%%%%%%%%%%%%%%%%%%%%%%%%%%%%%%%%%%%%%%%%%%%%%%%%%%
%%%%%%%%%%%%%%%%%%%%%%%%%%%%%%%%%%%%%%%%%%%%%%%%%%%%%%%%%%%%%%%%%%%%%%%%%%%%%%%

\end{document}